\PassOptionsToPackage{table,xcdraw}{xcolor}
\documentclass[acmsmall,screen]{acmart}
\usepackage{mathrsfs}
\usepackage{amsfonts}
\usepackage{algorithm}
\usepackage{algorithmicx}
\usepackage{algpseudocode}
\usepackage{graphicx} 
\usepackage{float} 
\usepackage{subfigure} 
\usepackage{multirow}
\usepackage{threeparttable}
\usepackage[table,xcdraw]{xcolor}
\usepackage{algpseudocode}
\usepackage{makecell}
\usepackage{colortbl}
\usepackage{bm}
\usepackage{xspace}
\usepackage[normalem]{ulem}
\usepackage{lineno}
\AtBeginDocument{%
  }

\setcopyright{acmlicensed}
\copyrightyear{2025}
\acmYear{2025}
\acmDOI{XXXXXXX.XXXXXXX}

\acmJournal{JACM}
\acmVolume{37}
\acmNumber{4}
\acmArticle{111}
\acmMonth{12}

\newcommand{\ie}{{\emph{i.e.}}\xspace}
\newcommand{\eg}{{\emph{e.g.}}\xspace}
\newcommand{\R}{}
\begin{document}
\title{DALD-PCAC: Density-Adaptive Learning Descriptor for Point Cloud Lossless Attribute Compression}

\author{Chunyang Fu}
\email{chunyang.fu@my.cityu.edu.hk}
\orcid{0000-0002-1487-9533}
\affiliation{
  \institution{City University of Hong Kong}
  \country{Hong Kong SAR, China}
}

\author{Ge Li}
\email{geli@ece.pku.edu.cn}
\affiliation{
  \institution{Peking University Shenzhen Graduate School}
  \city{Shenzhen}
  \country{China}
}

\author{Wei Gao}
\email{gaowei262@pku.edu.cn}
\affiliation{
  \institution{Peking University Shenzhen Graduate School}
  \city{Shenzhen}
  \country{China}
}

\author{Shiqi Wang}
\authornote{Corresponding author.}
\email{shiqwang@cityu.edu.hk}
\affiliation{
  \institution{City University of Hong Kong}
  \country{Hong Kong SAR, China}
}

\author{Zhu Li}
\email{zhu.li@ieee.org}
\affiliation{
  \institution{University of Missouri-Kansas City}
  \city{Kansas City, Missouri}
  \country{USA}
}

\author{Shan Liu}
\email{shanl@tencent.com}
\affiliation{
  \institution{Tencent Media Laboratory}
  \city{PaloAlto}
  \country{USA}
}


\begin{abstract}
  Recently, deep learning has significantly advanced the performance of point cloud geometry compression. However, the learning-based lossless attribute compression of point clouds with varying densities is under-explored. In this paper, we develop a learning-based framework, namely DALD-PCAC that leverages Levels of Detail (LoD) to tailor for point cloud lossless attribute compression. We develop a point-wise attention model using a permutation-invariant Transformer to tackle the challenges of sparsity and irregularity of point clouds during context modeling. We also propose a Density-Adaptive Learning Descriptor (DALD) capable of capturing structure and correlations among points across a large range of neighbors. In addition, we develop a prior-guided block partitioning to reduce the attribute variance within blocks and enhance the performance. Experiments on LiDAR and object point clouds show that DALD-PCAC achieves the state-of-the-art performance on most data. Our method boosts the compression performance and is robust to the varying densities of point clouds. Moreover, it guarantees a good trade-off between performance and complexity, exhibiting great potential in real-world applications. The source code is available at \href{https://github.com/zb12138/DALD_PCAC}{https://github.com/zb12138/DALD\_PCAC}.
\end{abstract}



\received{21-May-2024}
\received[revised]{27-Nov-2024}
\received[accepted]{01-Dec-2025}

\maketitle

\section{Introduction}
The point cloud is an unordered set of 3D points with six degrees of freedom, enabling realistic rendering of intricate scenes and objects. With the rapid development of 3D sensing and capture technology, point clouds are widely used in various applications such as virtual reality, unmanned aerial vehicles, surveying, immersive media, and self-driving cars~\cite{MUSCLE}. However, the massive influx of point cloud data from continuous scanning poses significant challenges in transmission and storage, especially in applications that require lossless transmissions, such as Geographic Information System (GIS), Computer-Aided Design (CAD), and cultural heritage protection~\cite{usecase}.
For example, an 18-bit LiDAR with a million points requires a bandwidth of 1.5 Gb/s if transmitted at 30 frames per second without compression. Meanwhile, the latency of the compression algorithms is crucial for real-time applications (\eg, automatic driving and immersive communication). In addition to geometry (\ie, XYZ), point clouds (PCs) usually associate various attributes such as colors (\ie, RGB), reflectance and normals. The bitstream of the attribute accounts for most of the total bitstream. Hence, it is necessary to explore the development of more efficient and effective algorithms for point cloud attribute compression (PCAC). In recent years, many methods for PCC have been developed, and most of them compress the geometry first and then compress the attribute based on the reconstructed geometry. Along this convention, this paper targets \R{lossless attribute compression on static point clouds}, assuming that the geometry of point clouds has been coded separately.

Compared with the long-standing compression problem of images and videos in regular grid forms, the challenge of PCC is that its geometry is sparse and irregular~\cite{quach2022survey}. This intricacy makes it difficult to directly apply 2D techniques to remove the 3D spatial and temporal correlations. Most compression methods overcome this problem by constructing or reorganizing the point cloud into regular forms, such as images, voxels, and octree. In this manner, video-based PCC (V-PCC) and geometry-based PCC (G-PCC) are two well-established  standards~\cite{graziosi2020overview,schwarz2018emerging} developed by Moving Picture Expert Group (MPEG). V-PCC projects 3D points into 2D images and compresses the patches employing traditional video coding techniques. G-PCC processes and compresses point clouds using octree for geometry compression, and techniques including region-adaptive hierarchical transform (RAHT)~\cite{de2016compression}, predicting transform, and lifting transform~\cite{gpcc} are proposed for attribute coding. In addition to traditional rule-based PCC solutions, learning-based PCC approaches have gained wide attention and significantly improved geometry compression efficiency~\cite{quach2022survey}. Yan {\em{et al.}}~\cite{yan2019deep} were the first to attempt to compress the geometry through autoencoder based on PointNet~\cite{qi2017pointnet}. Later, some voxel-based methods~\cite{VoxelDNNTCSVT,wang2022sparse,Guarda2021AdaptiveDL} and octree-based methods~\cite{fu2022octattention,EHEM,STAPCC} were proposed, which outperform G-PCC in both lossy and lossless geometry compression. \R{Recently, JPEG Pleno finalized a learning-based PCC standard~\cite{guarda2024jpeg} and developed a reference software Verification Model 4 (VM4) for lossy geometry and attribute coding of static point clouds.} 

\R{Compared to deep learning-based geometry compression methods, relatively few approaches focus on point cloud attribute compression, which may be due to the irregularity and sparsity of the attribute support, pose critical challenges for point cloud attribute compression~\cite{quach2022survey}. Deep-PCAC~\cite{Deep-PCAC} is a pioneer work for lossy attribute compression in the point domain, which uses second-order point convolution to exploit spatial correlations. Recent deep learning methods~\cite{fang20223dac, CNET, wangS} and the JPEG Pleno learning-based PCC standard~\cite{guarda2024jpeg} use sparse convolution for point cloud attribute coding. These methods mainly focus on the coding of dense point clouds and have achieved remarkable performance as dense point clouds retain original geometry information (\ie, planes, curves, lines) in local voxels with strong correlations, highlighting the pivotal role of geometry representation in attribute compression. However, most methods cannot adapt to point clouds with lower densities. Their effectiveness in sparse point clouds still falls short, as it is more challenging for convolutions to capture the features of neighboring points in a sparse scene.} 

\R{In sparse point clouds, such as LiDAR and large scene point clouds, the distance among the neighbors is often large (\eg, greater than ten voxels). }Voxelized convolution-based methods~\cite{CNET, wangS} are often incapable of capturing enough points due to the constraints imposed by the size of the convolution kernel. These issues limit their performance in the sparse case. \R{Although the sparsity problem can be mitigated through multi-scale and up-sampling}  approaches~\cite{wangS}, the context of an extensive range of neighbors is still underutilized. Moreover, limited by the structure of 3D convolution, the design of the autoregressive mode, such as voxel-wise autoregression\cite{CNET} or grouped autoregression~\cite{wangS}, could be inflexible. In addition, due to the network's computation capacity limitation, point clouds are usually divided into blocks before training. Current methods~\cite{nguyen2021multiscale, Deep-PCAC,10013714} use KD-tree or octree to perform block partitioning. These methods that only consider the continuity of geometry often lead to blocks with discontinuity and high variance of attributes, which may bring difficulties to normalizing the attributes within the block and will affect the generalization and performance of the network.

In summary, problems in existing learning-based methods for point cloud attribute compression can be summarized as 1) waste on memory and computation resources in voxel-based methods; 2) \R{limited receptive field for the context in sparse point clouds and thus cannot be applied to point clouds with different densities;} 3) the inflexible design of autoregressive mode limited by convolution structure, ultimately leading to slow decoding speed; and 4) inconsistent attributes with high variance within blocks, leading to unsatisfactory generalization capability.  Inspired by the success of 3D convolution on dense point clouds, we rethink the development of descriptors for point clouds. \R{To address these problems and further investigate a more efficient way for lossless attribute compression of static point clouds with different densities, we propose a point-based deep entropy model, which learns in the point space instead of the voxel space, and use the levels of detail (LoD) framework to build prior information. We propose a Density-Adaptive Learning Descriptor (DALD) to capture the structure and correlations among points searched in LoD over a wide expansion range, effectively accommodating point clouds with varying densities. We further developed a block partitioning method based on compressed attribute priors to reduce the attribute variance of blocks. Finally, we propose a deep entropy model utilizing a permutation-invariant Transformer for residual coding. Furthermore, }Our main contributions are summarized as follows:   
\begin{enumerate}
    \item[$\bullet$] We propose a point-wise attention framework based on a Transformer. The proposed entropy model is robust to point clouds with different densities, and achieves the state-of-the-art performance for lossless attribute compression.

    \item[$\bullet$] We develop a Density-Adaptive Learning Descriptor (DALD) to gather sufficient points for context modeling, thereby addressing the constraint of the receptive field inherent in traditional convolution-based methods, particularly for LiDAR and sparse point clouds.
   
    \item[$\bullet$] We devise a block partitioning method based on attribute priors to ensure the continuity of the geometry and attributes within blocks, thus reducing the attribute variance and improving the compression performance. 
    
\end{enumerate}

\section{Related work}
Point cloud compression involves geometry and attribute compression. In this section, we briefly review point cloud geometry compression and detail rule-based attribute compression and learning-based attribute compression.

\subsection{Point Cloud Geometry Compression}
The main idea to compress point cloud geometry is by coding the coordinates with organized structures (\eg, trees and images) and predicting their states. \R{In the lossless model, the occupancy states are coded by an algorithm coder using the predicted distribution. In the lossy model, the points (\ie, occupancy states) will be generated or approximated according to the predicted probabilities.} The traditional geometry compression methods in G-PCC~\cite{gpcc} include the prediction tree, the octree, and Trisoup. V-PCC ~\cite{graziosi2020overview} projects the point cloud into videos and compresses them using the video codec. \R{JPEG Pleno proposed a learning-based standard~\cite{guarda2024jpeg} upon an autoencoder for lossy coding of static point clouds.} Recently, a series of deep-learning based geometry compression works have also been developed \R{for both lossy and lossless compression}, which are mainly based on voxels~\cite{VoxelDNNTCSVT,wang2022sparse, Guarda2021AdaptiveDL}, octrees~\cite{fu2022octattention, EHEM,cui2023octformer, LoDhi2023sparse}, points~\cite{9354895,he2022density} and images~\cite{liu2023bird,ford2022lossless,zhou2022riddle}. In these methods, OctAttention~\cite{fu2022octattention} first proposed an attention-based method \R{for lossless geometry compression in octree domain}, which modeled the sibling and ancestor context in octrees and mitigated the problem of insufficient context for convolution. Attention-based methods~\cite{fu2022octattention, EHEM,cui2023octformer, ECM-OPCC} and Transformer-based works~\cite{Wang2023OctFormer, zhao2021point, lai2022stratified} also show excellent potential in point cloud compression and analysis tasks. This work provides evidence that using the Transformer in attribute compression tasks is also feasible.

\subsection{Rules-Based Point Cloud Attribute Compression}
There are numerous solutions for rule-based attribute compression, most based on transform, prediction, and projection. Methods based on geometry transformation form the mainstream of lossy attribute compression~\cite{xu2018cluster,huang2008generic,de2016compression,gu20193d}, exploiting attribute correlations between neighborhood points. For example, the graph Fourier transform (GFT) and its variants are extensively studied in~\cite{cohen2016attribute,shao2017attribute,xu2020predictive}, along with the exploration of Gaussian process transforms (GPTs) in~\cite{de2017transform}. \R{Some graph-based methods~\cite{watanabe2024fast,chen2024no} also play an important role in point cloud quality assessment and denoising}. However, the high computational cost of eigenvalue decomposition makes their methods inefficient. To alleviate this, researchers proposed a region-adaptive hierarchical transform (RAHT)~\cite{de2016compression} similar to a hierarchical subband-based adaptive Haar wavelet to better balance performance and complexity. Alongside enhancing RAHT using Set Partitioning~\cite{de2016compression} and B-spline wavelet basis~\cite{chou2019volumetric}, researchers have also integrated RAHT into the MPEG G-PCC reference software TMC13~\cite{MPEGGroup}, which has become a core prediction module similar to lifting transform~\cite{M42640} and prediction transform~\cite{N17249}. Prediction-based methods focus mainly on reordering the points and improving prediction performance. For example, G-PCC~\cite{gpcc} utilizes the levels of detail (LoD) structure that divides point clouds into multiple refinement levels, enabling prediction using varying detail levels. Furthermore, Chen {\em{et al.}}~\cite{chen2022hilbert} propose the Hilbert space-filling curve as an effective technique, and Yin {\em{et al.}}~\cite{yin2021lossless} introduce a normal-based predictor to improve the accuracy of the G-PCC prediction branch. \R{Hierarchical bit-wise differential coding (HBDC)} proposed in~\cite{huang2021hierarchical} uses bit-wise differential coding by exclusive-or operation to compress point cloud attributes in the octree framework. Methods based on projection aim to transform irregular 3D point clouds into the 2D regular form to apply existing image/video compression standards (\eg, JPEG and H.265)~\cite{mekuria2016design,li2020efficient}. The method introduced in~\cite{mekuria2016design} partitions the point cloud into 8×8 blocks, applies zigzag scanning to organize the attributes, projects them onto 2D images, and compresses using JPEG. Likewise, MPEG V-PCC ~\cite{graziosi2020overview} projects point cloud data into three associated videos (occupancy map, geometry video, and attribute video) encoded using 2D video coding standards. Our method adopts the prediction-based strategy using the LoD framework for lossless attribute compression.

\subsection{Learning-Based Point Cloud Attribute Compression}
Neural networks are still in the early age of point cloud attribute compression. \R{For lossy attribute compression}, the potential solution is to map irregular structures onto regular ones such that network architectures are capable of handling irregular input. Quach {\em{et al.}}~\cite{quach2020folding} train a lossy folding network to map 3D attributes to 2D grids and employ a video codec for compression. However, this collapse method introduces irreversible distortion and additional complexity. Another lossy compression approach is adopting the autoencoder framework that maps attributes to a 3D regular grid and uses 3D dense convolution to compress~\cite{alexiou2020towards}.
Analogously, Wang {\em{et al.}}~\cite{wang2022sparse} use efficient sparse convolution instead of 3D convolution, maps attributes represented by sparse tensors to the latent space, and encodes the latent vector using a hyperprior/autoregressive context model. Inspired by PointNet~\cite{qi2017pointnet}, Deep-PCAC~\cite{Deep-PCAC} proposes multilayer perceptrons (MLPs) in a point-based autoencoder, which can take point clouds directly as input. NF-PCAC~\cite{pinheiro2023nf} introduces a novel lossy attribute compression network based on normalization flow. \R{JPEG Pleno DL-PCC standard~\cite{guarda2024jpeg} projects and packs the recolored point clouds onto images and compresses them using an image codec. }\R{For lossless compression}, \R{MuSCLE~\cite{MUSCLE} is the earlier work that employs continuous convolution in the dynamic LiDAR intensity entropy model.} Recently, Wang {\textit{et al.}~\cite{wangS} proposed an autoencoder framework \R{3CAC} to predict the probabilities of attributes using cross-scale, cross-group, and cross-color correlations. Nguyen {\em{et al.}}~\cite{CNET} also propose a sparse convolution-based framework \R{CNeT} in both lossless geometry and lossless attribute compression, while the high computing overhead resulting from voxel-wise autoregression is still unacceptable. Although its fast version \R{MNeT}~\cite{MNet} reduces complexity through a multiscale structure, its performance is significantly affected by the lack of context. \R{These methods are usually weakest for sparse contents, as the convolution kernel cannot capture enough points to provide sufficient context.}

\section{The Proposed Method}
\subsection{Point Cloud Density}
\label{Point Cloud Density}
The density of a point cloud has a significant impact on the compression task in general. The point cloud density can be defined as the number of points per unit of space. Thus, when there are more points per unit voxel, point clouds tend to be denser. In such cases, point clouds typically exhibit smoother surfaces, with a stronger correlation between points, providing more contextual information. Because of the higher inter-point correlation, prediction and coding techniques can more accurately infer the uncoded points, resulting in higher compression ratios. However, some point clouds, such as architectural scenes, cultural heritage sites, and LiDAR, are often sparse. This sparsity can be attributed to several factors, including the large size of the scene, insufficient scanning, and \R{sampling}. Consequently, these point clouds have fewer points per voxel, resulting in a lower density. Sparse point clouds typically exhibit weaker inter-point correlations, and the distribution of points is more scattered. As a result, capturing sufficient contextual correlations becomes more challenging, consequently leading to lower compression ratios.  

\R{If we apply the convolution to voxelized point clouds, the unit space can be assumed to be the size of the convolution kernel. We propose a density metric, number of neighbors (NN), defined as the average number of points captured by the convolution kernel of predefined size (\eg, $5\times5\times5$) per point. Considering MPEG category 1 datasets~\cite{CTC} as an example, they primarily consist of static objects and scenes, and most point clouds are relatively sparse. For this dataset, the sorted NN are depicted in Figure~\ref{nnNum}, which represents the density (from sparse to dense) of the point cloud to some extent. To prove this, the MPEG standard~\cite{CTCchanges} also categorizes these point clouds into four categories: Scant, Sparse, Dense, and Solid, from ``very sparse and discontinuous'' to ``very dense with continuous surfaces''. This classification is consistent with the NN metric as shown in Figure~\ref{nnNum}. DALD-PCAC achieves state-of-the-art performance on point clouds that are not excessively dense, specifically including Scant, Sparse, and Dense\footnote{Capital beginning indicates the category names. The Dense may not deserve its name since it is still relatively sparse.} categories. For the Solid contents, our method still works and achieves a comparable performance to the traditional convolution-based methods~\cite{CNET,wangS}. However, these methods may struggle with the relative sparse point clouds, because the convolution kernels can only capture a few neighboring points as contexts (most are fewer than 10 points, as analysised in Figure~\ref{nnNum}).} In some worse cases, such as with Scant and Sparse contents, the traditional convolution kernels may not capture any neighbors. Consequently, the loss of contextual information poses a challenge for mainstream convolution-based methods to model inter-point correlations, resulting in subpar performance. 

Based on this observation, \R{we propose a point-based compression framework to tackle the challenges of sparsity and irregularity in lossless attribute compression of point clouds. We utilized a Levels of Detail (LoD) structure to obtain prediction residuals and construct a sufficient context for each point. Additionally, we design a Density-Adaptive Learning Descriptor (DALD) for embedding local neighbors, enhancing inter-point correlations' modeling capability. Finally, we present a Transformer-based entropy model for residual coding, significantly improving the performance of lossless attribute compression.}

\begin{figure}[t]
\centering
{\includegraphics[scale=0.33]{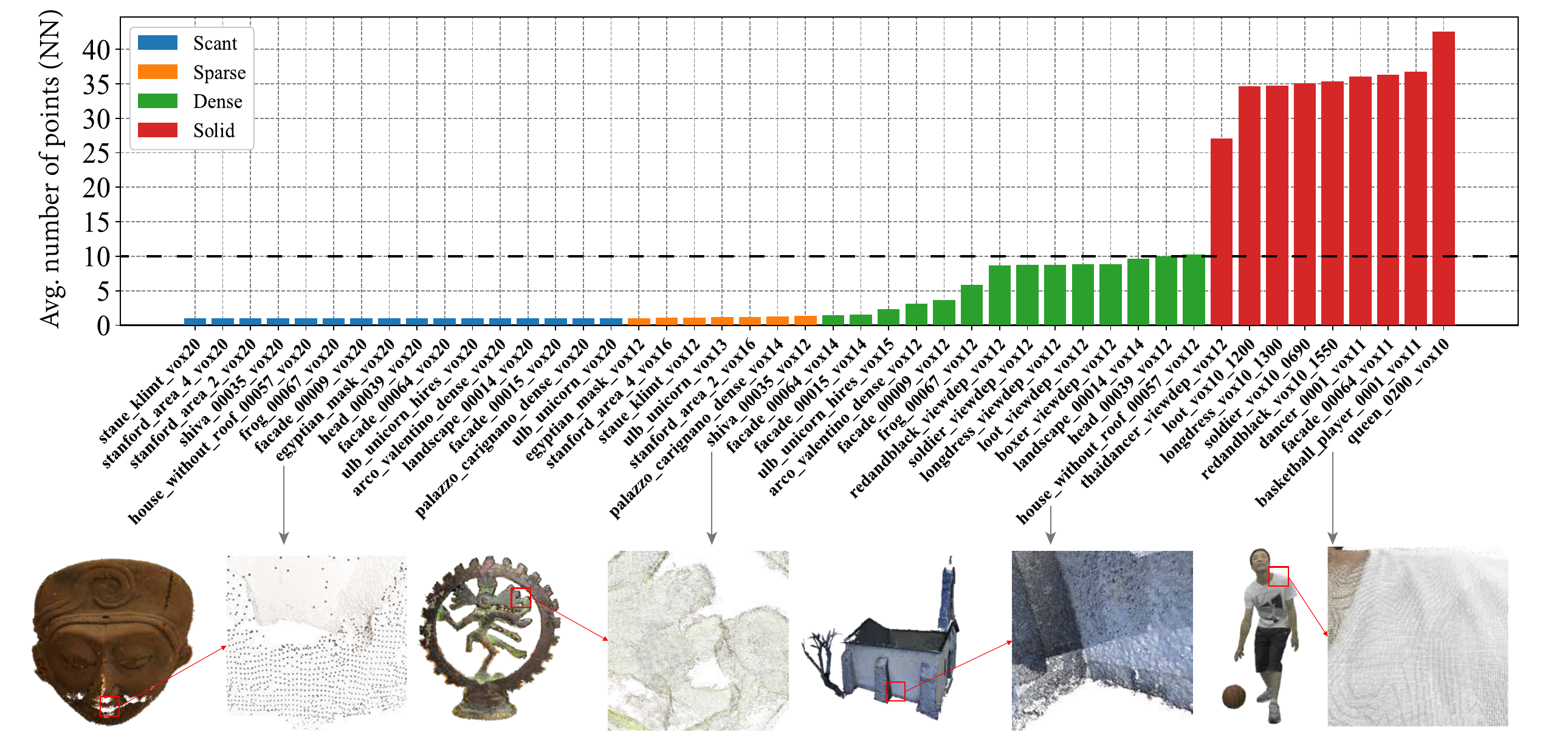}}
\caption{The average of the number of neighbors (NN) that are captured by a $5\times5\times5$ convolution kernel for different point clouds. The number of neighbors is less than 10 points in most sparse point clouds.}
\label{nnNum}
\end{figure}

\subsection{Overview}
In this paper, we focus only on lossless compression of the attribute. Therefore, as in many related works, we assume that the geometry (\ie, coordinates) of the point clouds has been transmitted and reconstructed. \R{Specifically, in the case of lossy geometry reconstruction, the discussed method still applies to the lossless compression of the recolored attributes.} The overview of our method is illustrated in Figure~\ref{overview}. Inspired by the predicting transform in G-PCC~\cite{gpcc}, we introduce two methods to generate Levels of Detail (LoD). LoD aims to divide the point cloud into multiple disjoint point sets, called \textit{refinement layers}. \R{First, the distance-based LoD generation partitions the input point cloud into $T$ refinement layers, denoted as $R_1, R_2,..., R_{T}$ and some remaining points. The refinement layers $R_1, R_2,..., R_{T}$ collectively form the \textit{Base Layer}, initially representing the point cloud. Second, we propose a sampling-based method to divide the remaining points into additional $L-T$ refinement layers $R_{T+1},..., R_{L}$, and together they constitute the \textit{Inference Layer}.} $LoD_i$ is defined as the union of the current $R_{i}$ and previous refinement layers, that is, $R_1\cup R_2\cup...\cup R_{i}$.  

\R{The data flow of our proposed compression framework is depicted in Figure~\ref{Flowchart}. After LoD generation, the input point cloud is split into Base Layer and Inference Layer.} The Base Layer has fewer points and sparser and thus it is more difficult to compress. Therefore, we employ both inter-layer and intra-layer predictions, which utilize both neighboring points from the previous LoD and preorder points in the current refinement layer to predict the attributes. The prediction residuals are subsequently compressed using Run-length coding~\cite{Run_length} directly. The Inference Layer consists of a larger proportion of points and thus we exclusively utilize inter-layer prediction to avoid the fully autoregressive problem. In order to achieve parallelism under the constraint of GPU memory, we propose a prior-guided block partitioning method, based on the Base Layer to divide the Inference Layer into blocks with lower variance. Those blocks are embeded by DALD and are compressed by a point-wise attention entropy model. The entropy model predicts the attribute residual distributions for all the points in parallel, and the distributions will guide the arithmetic coder to perform lossless compression and generate the bitstream. At the decoder, the Base Layer is decoded first, and the successive refinement layers of Inference Layer are decoded by blocks. 

\begin{figure*}[t]\centering
	\includegraphics[scale=0.4]{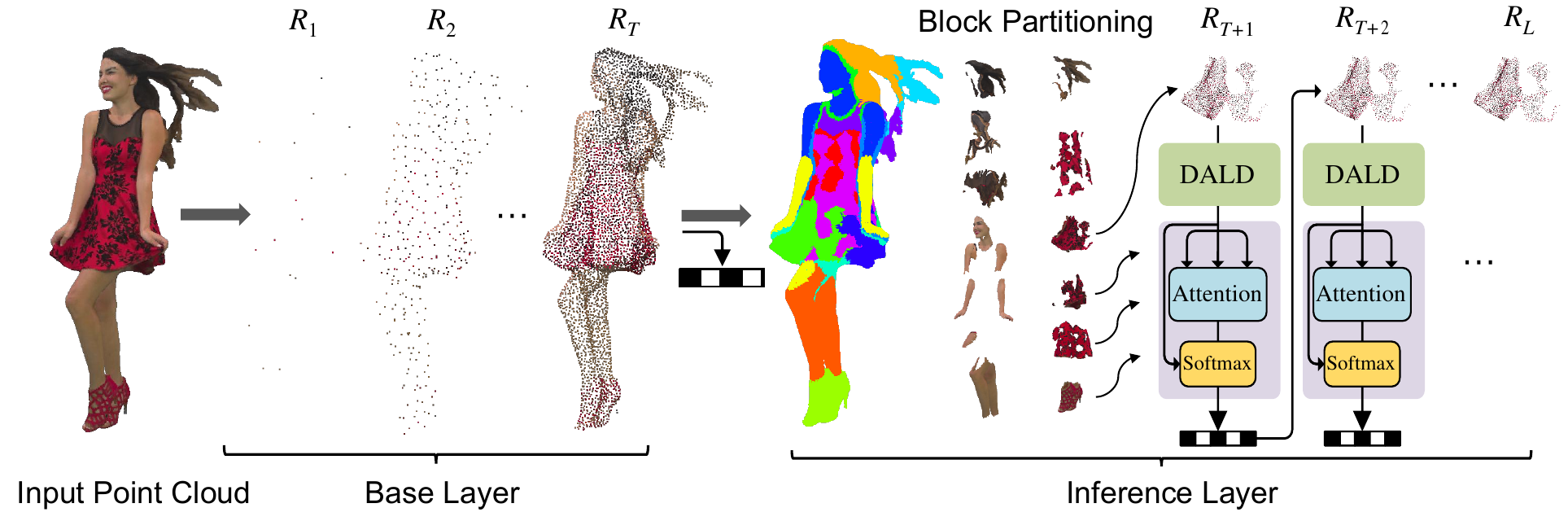}
	\caption{LoD architecture of our method. Input point cloud is sampled into \textit{Base Layer} and \textit{Inference Layer} which are coded separately. ${R_1,...,R_L}$ represents refinement layers. ``DALD'' represents Density-Adaptive Learning Descriptor (Sec. \ref{DALD}). Blocks are compressed by the Deep Entropy Model (Sec. \ref{DALD}) in parallel.}
    \label{overview}
\end{figure*}

\begin{figure}[t]
\centering
{\includegraphics[scale=0.5]{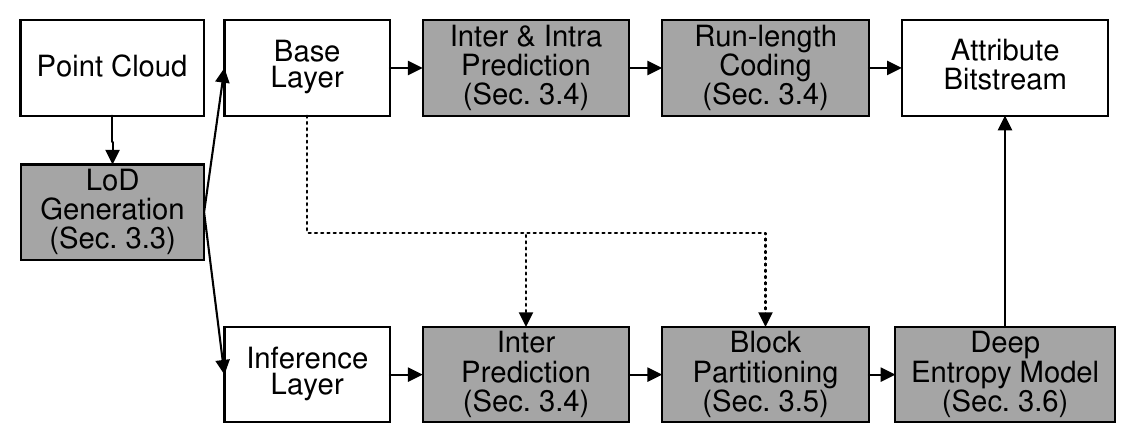}
\caption{\R{Flowchart of the proposed attribute compression framework. Dashed lines represent providing context.}}
\label{Flowchart}
}
\end{figure}
\begin{figure}[htb]
\centering
\includegraphics[scale=0.5]{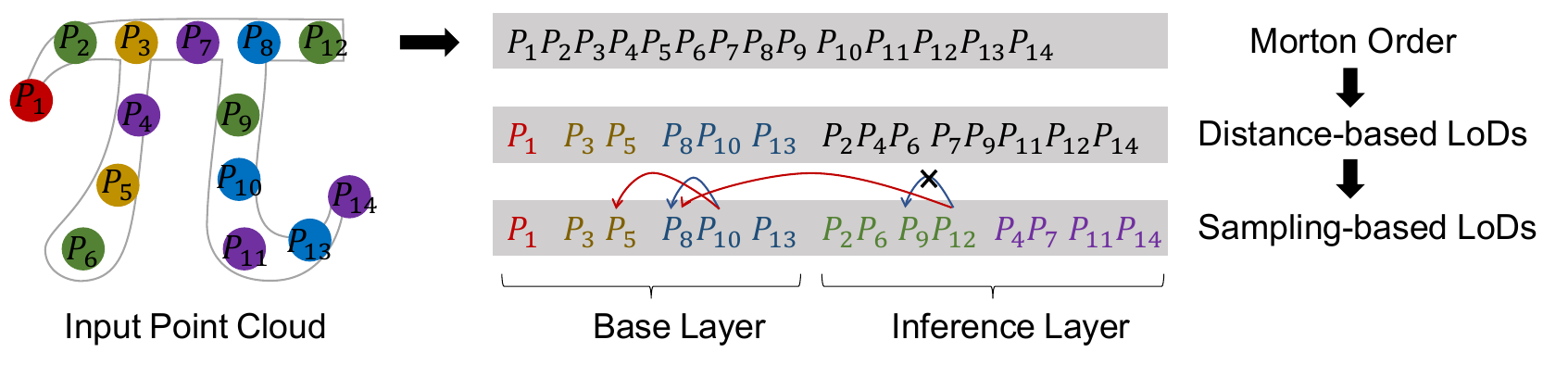}
\caption{\R{LoD generation example. Points of each color form a refinement layer. In the Inference Layer, intra-prediction (blue arrow) is disabled.}}
\label{algorithm1}
\end{figure}

\subsection{Levels of Detail Generation}
There are many ways to generate LoDs and search for $K$ nearest neighbors (KNN), including G-PCC~\cite{gpcc} and Wei {\em{et al.}} method ~\cite{LoD}. \R{We propose a hybrid LoDs generation method to balance the complexity and prediction performance}. \R{An example of this hybrid LoDs is described in Figure~\ref{algorithm1}. The distance-based levels of detail generation, same as G-PCC~\cite{gpcc}, is first employed to generate the Base Layer including $T=3$ refinement layers (\ie, $\{P_1\},\{P_3,P_5\},\{P_8,P_{10},P_{13}\}$). In this process, neighbors from both the previous refinement layers (\ie, inter, red arrow) and current refinement layers (\ie, intra, blue arrow) are searched to achieve a more accurate prediction, For example, points $P_{5}$ and $P_{8}$ are used to predict $P_{10}$. The remaining points $\{P_2,P_4,P_6,...,P_{14}\}$ are even divided into $L-T=2$ refinement layers by uniformly spaced sampling and form the Inference Layer. The intra-prediction in the Inference Layer is disabled (\eg, $P_9$ predicts $P_{12}$) because it will cause the autoregressive problem that increases decoding time. Two main reasons exist for dividing these points into additional layers through uniform sampling. First, this approach allows us to evenly maximize the number of points in each refinement layer with low complexity to reach the bound of GPU memory, thereby maximizing the parallelism, as a refinement layer can be encoded and decoded simultaneously during a single network inference. Second, since disabling intra-layer prediction has somewhat compromised prediction performance, increasing the number of refinement layers can supplement inter-layer prediction, thereby enhancing compression performance. Overall, the number of refinement layers in the Base Layer and Inference Layer is a trade-off between compression performance and complexity.} 

\subsection{Prediction}
\label{Prediction}
Prediction aims to reduce the residuals and obtain a more concentrated distribution of residuals. \R{The lossless compression of attributes is equivalent to lossless compression of the residuals.} $k$ nearest neighbors $\{p_j|j\in \mathcal{N}_i\}$ for each point $p_i$ are searched by KD-tree while in the above LoD construction, and them are used to draw prediction residuals $r_i = a_i - \hat{a}_i$ through inverse distance weighting (IDW) interpolation,
\begin{equation}
\label{eq1}
\hat{a}_i= \operatorname{round}\left(\sum_{j\in \mathcal{N}_i} \frac{{\frac{1}{d_{ij}^2}a_j}}{\sum_{k\in \mathcal{N}_i} \frac{1}{d_{ik}^2}}\right),
\end{equation}
where $a_i$ is the attribute of point $p_i$ and $d_{ij}$ is the Manhattan distance between $p_i$ and $p_j$. The neighbor searching is performed during the LoD construction to save time. Although the Inference Layer only uses the inter-layer prediction, it often demonstrates better prediction due to the availability of a sufficient number of closely related neighboring points.

The residuals of the Base Layer are directly compressed by run-length coding~\cite{Run_length} following the arithmetic coder. The residuals of the Inference Layer are compressed using a deep entropy model. Furthermore, these searched neighbors are used as the context in the deep entropy model.

\subsection{Block Partitioning}
\label{Block Partitioning}
\begin{figure}[t]
\centering
\subfigure[Block partitioning using KD-tree.]{
\includegraphics[height=0.33\textwidth]{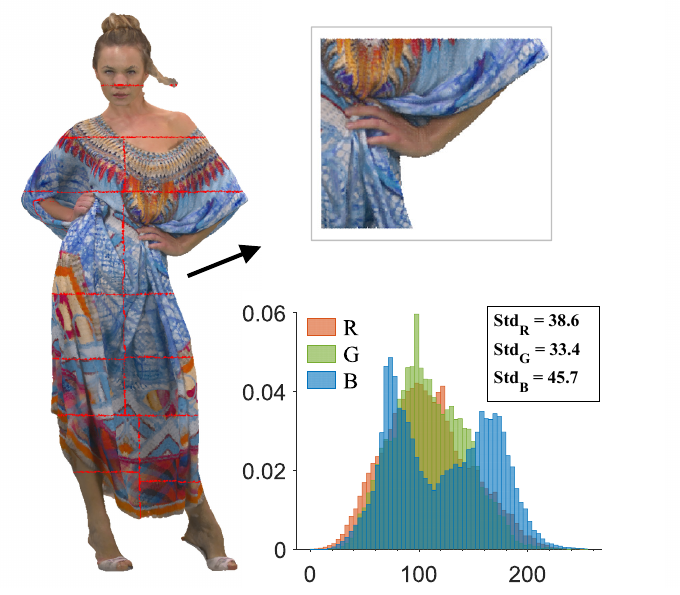}
}
\quad
\subfigure[Block partitioning using base-layer priors.]{
\includegraphics[height=0.33\textwidth]{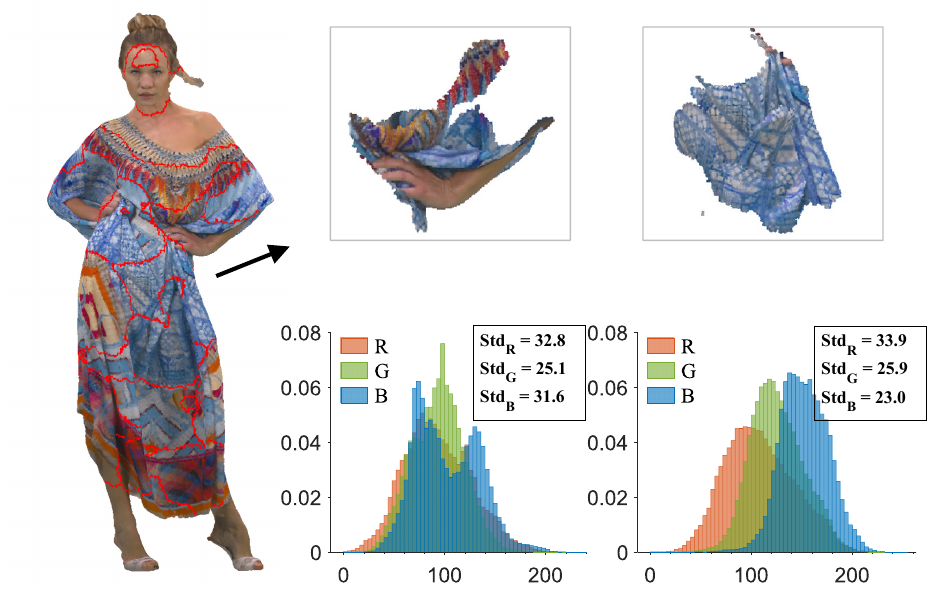}
}
\caption{Comparison of block partitioning methods based on KD-tree and the proposed Base Layer priors.}
\label{blocks1}
\end{figure}
\R{Since the Transformer-based entropy model requires $\mathcal{O}(N^2)$ memory space,} the refinement layer contains too many points for the model to process. Therefore, we have to divide it into blocks to improve the parallelism of the model. We expect the distribution of each block to be as consistent as possible to improve the robustness of the network. Previous methods~\cite{nguyen2021multiscale, Deep-PCAC,10013714} used an octree or a KD-tree to divide the entire point cloud into many blocks and then individually process them. However, these methods may not be helpful for attribute compression because they only consider the continuity of geometry, ignoring the consistency of attributes, which also makes the normalization difficult. As an example shown in Figure~\ref{blocks1} (a), in the KD-tree-based block partitioning, there are many boundaries of discontinuity of the attributes, leading to a high variance of the attributes in a block. 
To address this issue, we propose a partitioning method guided by the priors derived from the reconstructed Base Layer, which is compressed and reconstructed before. This method contains the following processes: 

First, the points in the Base Layer are clustered by the K-means algorithm base on the joint features of geometry and attributes,
\begin{equation}
    bkid_{i} = \operatorname{K-means}(\operatorname{concat}[\tilde{a}_i,\alpha\tilde{p}_i]),\ p_i\in \operatorname{Base\ Layer},
\end{equation}
where $\tilde{a}_i$ is the smooth attribute obtained by average filtering with a neighbor radius of 50 points, $\tilde{p}_i$ is the min-max normalized coordinate in $[0, 1]^3$, and $\alpha$ is the factor that controls geometry continuity of the clustering result, which is set to 255 empirically. The number of clusters is determined by the ratio of number of points in Inference Layer to the predefined size of blocks, which is generally $16\sim64$. Then, the block index of each point $p_j$ in the Inference Layer is set to same as its closest point belonging to the Base Layer,
\begin{equation}
    bkid_{j} = bkid_{i}, i = \operatorname{argmin}_i{\Vert p_i-p_j \Vert_2},\ p_i\in \operatorname{Base\ Layer}, \ p_j\in \operatorname{Inference\ Layer}.
\end{equation}
\R{Finally, points assigned the same $bkid$ are grouped into a block. Subsequently, every $N$ points that make up a batch are fetched from each block in Morton order, with padding applied as necessary.} 

\R{The partitioning results are shown in Figure~\ref{blocks1} (b). The proposed method usually has a more consistent distribution and the average variance is smaller than the counterparts in KD-tree-based methods. In contrast to other methods \cite{nguyen2021multiscale, Deep-PCAC,10013714} that leave the boundary points lacking of reference, the block partitioning in LoD offers another advantage.} The points at the block boundaries can still refer to adjacent blocks belonging to the previous LoD, because the correlations between blocks are maintained due to the KNN searching being performed before the block partitioning (as discussed in Section \ref{Prediction}). 

\subsection{Deep Entropy Model}
\label{DALD}
\R{As shown in Figure~\ref{motivation} (a), the traditional convolution's receptive field is fixed and limited when handling point clouds of varying densities. It may fail to capture neighbors as context when the point cloud becomes sparse. This explains why convolution-based methods are sensitive to the density of point clouds. To address this issue, we leverage the fixed number of neighbors $\{p_j\}$ from the perdition process (Section.~\ref{Prediction}) in LoD as context. In place of convolution, we propose a \textbf{Density-Adaptive Learning Descriptor} (DALD) for local feature description. DALD contains neighbors embedding and central point embedding $E_l(\cdot)$, $E_a(\cdot)$ and $E_r(\cdot)$ as learnable operations.}

\begin{figure}[t]
\centering
\subfigure[traditional Convolution]{
{\includegraphics[scale=0.425]{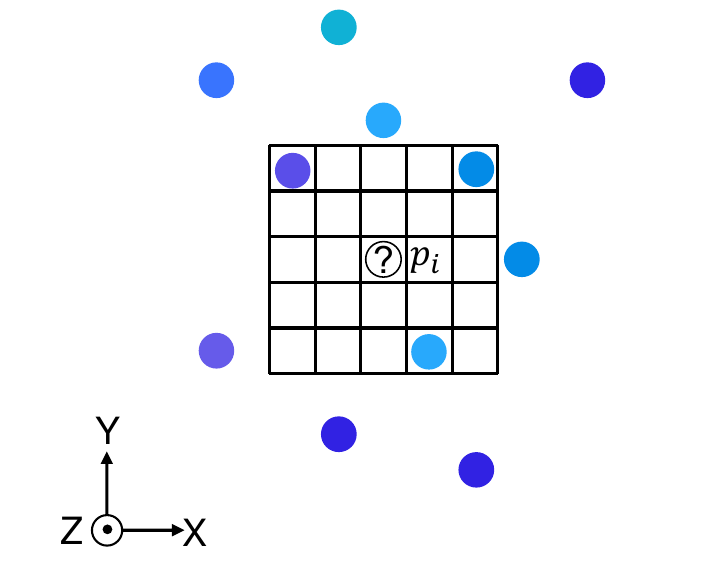}}
}
\subfigure[proposed  Density-Adaptive Learning
Descriptor]{
{\includegraphics[scale=0.425]{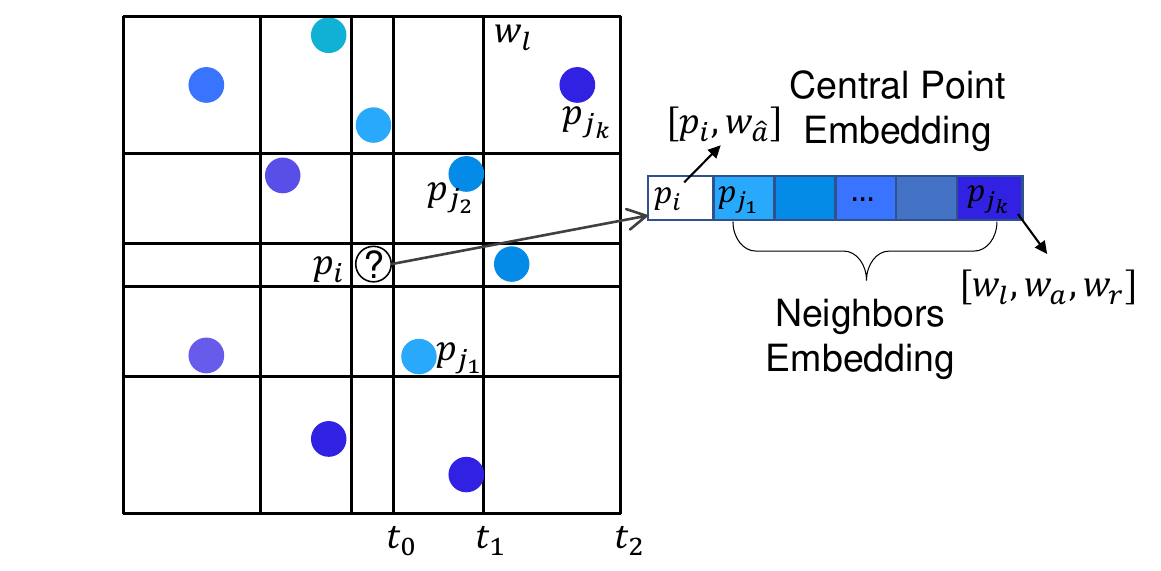}}
}
\caption{Motivation illustration of  Density-Adaptive Learning
Descriptor (DALD) when encoding $p_i$. (a) Transitional convolution with kernel size $5\times5\times5$;  \R{(b) Proposed DALD with $n=2$. The colors represent the attributes. $\{p_j\}$ are the searched neighbors of $p_{i}$ in the previous LoD. Only visualization of the XY plane is for simplicity.}}
\label{motivation}
\end{figure}

In the \textbf{neighbors embedding}, as shown in Figure~\ref{motivation} (b), we propose a Relative Position Embedding to embed the relative position (\ie, orientation and distance) of neighbors. It first embeds the relative position on each axis independently and then combines them into a final label. Herein, we take the X axis as an illustration. For each neighbor $p_j$ searched in LoD, we generate label $l_x\in[0,2n]$ by the distance between $p_j$ and central point $p_i$ along the X axis:
\begin{equation}
\begin{split}
l_{x} = \operatorname{sign}(x_j-x_i)*k + n\\ 
\text{ if}\ t^x_{k-1}<\frac{\operatorname{abs}(x_j-x_i)}{\bar{d_x}}\leq t^x_{k},
\end{split}
\label{Eq2}
\end{equation}
where $0\leq k\leq n$, $\bar{d}_x=\sum_i{\operatorname{abs}(x_{j_1}-x_i})/N$ is the average distance between the nearest neighbors and central points along the X axis in the current batch, and $\{t^x_i\}=t^x_{-1}, t^x_0, t^x_1, ..., t^x_n$ are predefined increasing thresholds (suppose $t^x_{-1}<0$ and $t^x_n=\infty$). In general, points are classified into $2n+1$ labels along the X dimension. Concerning the relative position embedding in 3D space, the final label of $p_j$ is the combination of the labels of each axis:
\begin{equation}
l = l_{x} + l_{y}*(2n+1) +l_{z}*(2n+1)^2.
\label{eq3}
\end{equation}
Herein, $\{t_n^x,t_n^y,t_n^z\}$ are set to $\infty$ in our experiment. Thus, we can gather neighbors into $(2n+1)^3$ kernels in all orientations and at all distances. In analogy to convolution kernels, we use an embedding operation to allocate different parameters for points with different labels: $E_l(l)=w_{l}$, where $l=0,1,...,(2n+1)^3-1$, and $w_{l}$ constitute the trainable parameter $W \in \mathbb{R}^{(2n+1)^3\times\cdot}$ of the embedding operation. In summary, we assign different learnable parameters to all the neighbors of points by bins similar to convolution kernels without worrying about the varying densities. 

Besides the relative location, we also believe that the attribute residuals (\ie, $a_i - \hat{a}_i$) of central point are also associated with relative attribute $a_j - \hat{a}_i$ and the ground truth $a_j - \hat{a}_j$ of its neighbors. Hence, those features are also added to the local descriptors as supplements, as shown in Eqn.~\eqref{Eq6}. For the robustness of the deep entropy model, $p_i$ are mapped to [0,1] over the batch by Min-Max normalization and $\hat{a}_i$, $a_j - \hat{a}_i$ and $a_j - \hat{a}_j$ are normalized by embedding when they constitute the final features. Formally, given a set of coordinates $\{p_i\} \in \mathbb{R}^{N\times3}$ in the current batch, each point has $k$ neighbors $\mathcal{N}_i=\{j | j_1,...,j_k\}$ from the last LoD. For its each neighbor, we set:
\begin{equation}
\bm{f}'_j=\operatorname{concat}[E_l(l_j),E_a(a_j),E_r(a_j - \hat{a}_i)],
\label{Eq6}
\end{equation}
as the initial features of neighbor $j$, where $E_l(\cdot)$, $E_a(\cdot)$ and $E_r(\cdot)$ are embedding operations.

In the \textbf{central point embedding}, we extract the initial feature of the central point $i$ as:
\begin{equation}
\bm{f}_i=\operatorname{concat}[p_i,E_a(\hat{a}_i)],
\label{Eq5}
\end{equation}
where $p_i$ is the location of point $i$ and $\hat{a}_i$ is the prediction value drawn from Eqn.~\eqref{eq1}.

We concat $\bm{f}_i$ and features $\bm{f}'_{j}$ of its $k$ neighbors as a local descriptor of point $i$:
\begin{equation}
\bm{g}_i=\operatorname{concat}[\bm{f}_i,\bm{f}'_{j_1},...,\bm{f}'_{j_k}].
\end{equation}
Each point has $k$ neighbors that can be used as context to predict the distribution of its attribute residuals currently. According to the theory that conditioning reduces entropy~\cite{el2011network}, we can expand the context to the whole batch to reduce the uncertainty of the attribute. Considering that points with similar features have similar attribute residuals, we use a point-wise attention network to find this similarity. Specifically, we set the aggregate contexts of points $p_1,...,p_N$ as follows,
\begin{equation}
\bm{C}_{1},\bm{C}_{2}...,\bm{C}_{N}= \bm{F}(\bm{g}_1,\bm{g}_2,...,\bm{g}_{N}),
\label{eq9}
\end{equation}
where $\bm{F}$ is an attention-based network and it predicts contexts for points in a batch simultaneously. 
Noting that there are no reference among points $p_1,...,p_N$, therefore there is no need to incorporate any mask operations, which eliminates autoregression in refinement layers and greatly improves the decoding speed. We use a simple yet effective three-layer Transformer encoder\footnote{https://pytorch.org/docs/stable/generated/torch.nn.Transformer.html}~\cite{vaswani2017attention} to model $\bm{F}$ without bells and whistles. The input and output dimension of this Transformer network is the dimension of $g_i$, \ie, $k\times(N_{E_l}+N_{E_a}+N_{E_r})+3+N_{E_a}$, where $N_{E}$ is the feature dimension after embedding operation $E$, and $3+N_{E_a}$ is the dimension of $\bm{f}_i$. In this way, we can encode the whole batch in one forward. It is worth mentioning that the Transformer is a permutation-invariant backbone~\cite{guo2021pct}, such that we do not have to specifically consider the ordering of the input $g_i$.

Set $Q(\bm{r})$ as the joint probability distribution of attribute residuals to be encoded in the current batch, and factorize this joint distribution into a product of conditional probabilities as our entropy model:
\begin{equation}
\label{factor}
Q(\bm{r})=\prod_i q\left(r_i \mid \bm{C}_i; \mathbf{w}\right),
\end{equation}
where $r_i= a_i - \hat{a}_i$ is the attribute residuals and $\mathbf{w}$ denotes the context model parameters. Suppose that the single attribute (\eg, reflectance, and intensity) is in 8 bit integers, and thus the residuals are in $[-255,255]$. \R{We make a 511-way prediction through a final MLPs with softmax,
\begin{equation}
q_i\left(r_i \mid \bm{C}\right)=\operatorname{SoftMax}\left(\operatorname{MLP}\left(\bm{C}_{{i}}\right)\right).
\label{sofmax}
\end{equation}
According to Shannon’s source coding theorem~\cite{el2011network}, the estimated bitrate of the residuals is cross-entropy $\mathbb{E}_{\bm{r} \sim P}\left[-\log _2 Q(\bm{r})\right]$, therefore, we minimize the cross-entropy between the predicted distribution and the ground truth. The loss function is,
\begin{equation}
\ell=-\frac{1}{N}\sum _i ^ N\log _2 q_i\left(r_i \mid \boldsymbol{C}\right).
\label{CEOneChannel}
\end{equation}}

\subsection{Multi-channel Attributes Compression}
\label{crossChannel}
More point clouds have more than one channel of attributes (\eg, RGB). We propose a multi-channel attributes compression module. To better remove the correlation among channels, RGB is converted into Y-CoCg-R~\cite{malvar2003ycocg} format losslessly,
\R{\begin{equation}
\begin{aligned}
    Co & = R - B \\
    t & = B + (Co \gg 1) \\
    Cg & = G - t \\
    Y & = t + (Cg \gg 1) \\
\end{aligned} \quad \Longleftrightarrow \quad 
\begin{aligned}
    t & = Y - (Cg \gg 1) \\
    G & = Co + t \\
    B & = t - (Co \gg 1) \\
    R & = B + Co
\end{aligned}.
\end{equation}}
Herein, brightness Y is in 8-bit integers and chrominance Co, Cg are in 9-bit integers. The residuals of Y, Co and Cg can be drawn from Eqn.~\eqref{eq1}, respectively. Assuming the residuals are $r_Y, r_U, r_V$, the joint probability of multi-channel attribute residuals can be decomposed into:
\begin{equation}
Q(r_Y,r_U,r_V) = Q_Y(r_Y)Q_U(r_U|r_Y)Q_V(r_V|r_Y,r_U).
\end{equation}
Because the ranges of the residuals are different, we truncate the residuals of Co, Cg into $[-255, 255]$ to remain consistent with Y. \R{After factorization following Eqn.~\eqref{factor}, similar to Eqn.~\eqref{sofmax}, we use the 511-way softmax to model $q(r_Y)$, and use simple multi-layer perceptrons (MLPs) to continue modeling $q(r_U)$ and $q(r_V)$,}
\begin{equation}
\label{eq10}
\begin{aligned}
& q_i\left(r_{Y} \mid \bm{C}\right)=\operatorname{SoftMax}\left(\operatorname{MLP}_Y\left(\bm{C}_{{i}}\right)\right),\\
& q_i\left(r_{U} \mid \bm{C}, r_{Y_i}\right)=\operatorname{SoftMax}\left(\operatorname{MLP}_{U}\left(\left[\bm{C}_{{i}}, r_{Y_i}\right]\right)\right), \\
& q_i\left(r_{V} \mid \bm{C}, r_{Y}, r_{U}\right)=\operatorname{SoftMax}\left(\operatorname{MLP}_{V}\left(\left[\bm{C}_{{i}}, r_{Y_i}, r_{U_i}\right]\right)\right).
\end{aligned}
\end{equation}
Run-length coding~\cite{Run_length} is used to encode the values outside the truncated range,  with the bitrate denoted as $R_{\text{out}}$. An advantage of modeling inter-channel dependencies in this way is that, there is little additional coding time, because the time consumption is mainly from the calculation of $\bm{C}_i$ in Eqn.~\eqref{eq9}. Accordingly, loss function Eqn.~\eqref{CEOneChannel} is modified to,
\begin{equation}
\ell=-\frac{1}{N}\sum _i ^ N\left[\log _2 q_i\left(r_{Y} \mid \boldsymbol{C}\right)+\log _2 q_i\left(r_{U} \mid \boldsymbol{C,r_Y}\right)+\log _2 q_i\left(r_{V} \mid \boldsymbol{C,r_Y,r_U}\right)\right]. 
\end{equation}
$N$ is the number of points in batches. The theoretical bitrate for the entire point cloud is
\R{\begin{equation}
R=\frac{L_{\text{base}}+(\ell+R_{\text{out}})*N_{\text{infer}}}{N_{\text{base}}+N_{\text{infer}}},
\label{Eq14}
\end{equation}
where $N_{\text{base}}$ and $N_{\text{infer}}$ represent the numbers of points in Base Layer and Inference Layer respectively. $L_{\text{base}}$ is the length of the Base Layer bitstream encoded by the run-length coding, and usually takes less than 20\% of the final bitstream.  $R_{\text{out}}$ is the rate of coding the truncation of the Co, Cg channels, and usually takes less than $10^{-4}$ bpp. $\ell$ is the Shannon entropy of the residuals of the Inference Layer, which is the optimization objective of the deep entropy network.}
 
\section{Experiments}
\subsection{Datasets}
    Following the method of CNeT~\cite{CNET} and Wang {\em{et al.}}'s method, (\emph{a.k.a} 3CAC)~\cite{wangS}, \R{to make a fair comparison,} we use Andrew, David and Sarah in MVUB~\cite{MVUB}, Longdress and Soldier in 8iVFBv2~\cite{8iv2} and Owlii~\cite{Owlii} for training (3882 point clouds in total, denoted as \textit{CNeT Training Set}, see Figure~\ref{dataset}). \R{To further verify the performance of our method across datasets, we also train our model on a larger colored point cloud dataset PCL-PCD~\cite{pcd-pcl}, which consists of 80 PCs of large objects sampled from Sketchfab~\footnote{https://sketchfab.com/} meshes. Generally, PCL-PCD is sparser, with NN ranging from 1.00 to 7.12, and has a rich diversity in texture.} Following the recommendations of the MPEG AI-based point cloud experiments~\cite{AIPCC}, we evaluated our model using the PCs suggested from the MPEG CAT1 dataset~\cite{CTC}. This dataset includes point clouds with varying densities, ranging from Scant to Solid. The density of these data is illustrated in the proposal~\cite{CTCchanges}, \R{which is summarized in Table~\ref{table1}, and it can also be validated by the NN value proposed in the Figure~\ref{nnNum}}. We also test our model on some dense point clouds (\ie, 8iVFBv2 and MVUB) to verify the density generalization of our method.  

Our method is also applicable to LiDAR point clouds, which are a type of sparse point clouds. The density of LiDAR is comparable to the Sparse types in Figure~\ref{nnNum}. Following 3CAC, we performed experiments on LiDAR datasets that included Ford~\cite{pandey2011ford} ($3\times1500$ scans) and SemanticKITTI~\cite{behley2019semantickitti} sequences (43552 scans). We use KITTI sequences 00-10 and Ford\_01\_q1mm for training, frames ended with `00' in Ford\_03\_q1mm for validation and others for testing. Since SemanticKITTI is in floats, we quantize its geometry into $12\sim16$ bits over a $[0,400\text{m}]^3$ region and quantize the intensity into 8 bits following MuSCLE~\cite{MUSCLE}. 

\begin{figure*}[t]\centering
	\includegraphics[width=\linewidth]{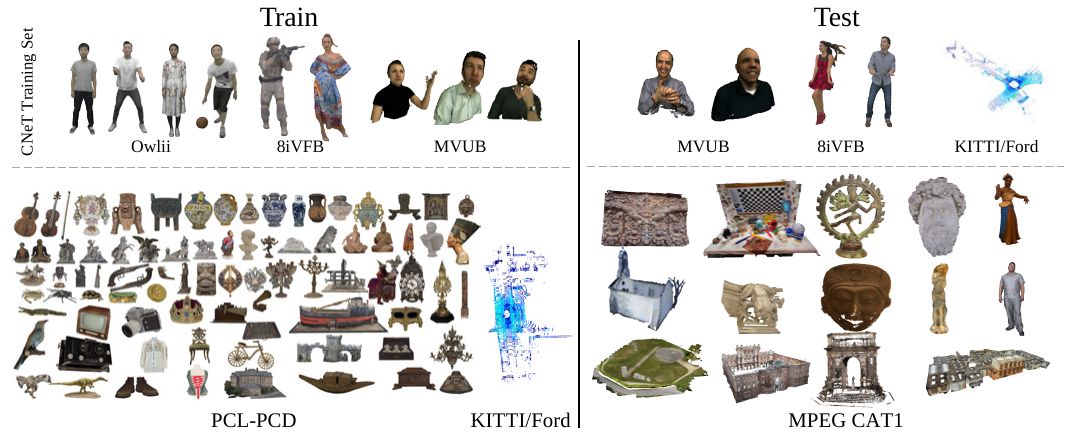}
	\caption{\R{Datasets used in this research.}}
    \label{dataset}
\end{figure*}

\begin{table}[]
\centering
\caption{Experiment Setting in Object Point Clouds and LiDAR Compression}
\label{Parameters}
\begin{tabular}{c|c|c}
\hline 
                        & Object Point Cloud& LiDAR                    \\ \hline \hline
Attribute               & Three 8-bit colors & Single 8-bit reflectance \\  
LoD                     & $T=8\ L=24$        & $T=8\ L=16$                         \\  
Points in Batch           & $N=1024$               & $N=4096$                     \\  
Number of Neighbors& $k=11$              & $k=9$                      \\ \hline
DALD Thresholds (n=3)& $\{t^{x}_i\}=\{t^{y}_i\}=\{t^{z}_i\}=\{0,1,3,\infty\}$   &  \makecell[l]{$\{t^x_i\}=\{t^y_i\}=\{0.2,1,3,\infty\}$ \\ $\{t^z_i\}=\{0.2,0.4,1,\infty\}$} \\\hline
$E_l$, $E_a$, $E_r$     &  $\mathbb{R}^{343\times6}$, \makecell[l]{$\mathbb{R}^{256\times3}, \mathbb{R}^{511\times6}\ $ for Y\\$\mathbb{R}^{512\times3}, \mathbb{R}^{1023\times6}$ for Co/Cg }  &$\mathbb{R}^{343\times6}, \mathbb{R}^{256\times3}, \mathbb{R}^{511\times6}$  \\ \hline
Training Dataset           & \makecell[c]{\R{CNeT Training Set / PCL-PCD Set}}  &   \makecell[c]{Ford\_01 \\ SemanticKITTI 00-10}  \\\hline
Testing Dataset            & \makecell[c]{Ricardo, Phil \\ Loot, Redandblack \\ MPEG CAT1 Set} &   \makecell[c]{Ford\_02,03 \\ SemanticKITTI 11-21}                   \\\hline 
\end{tabular}
\end{table}
\subsection{Model Implementation and Baselines}
We set some parameters to our model for point clouds of objects and LiDAR. In object point clouds, we construct $T=8$ levels of refinement layers according to the default setting in G-PCC~\cite{gpcc}, and the remaining points are evenly divided into $L-T=16$ layers that make up the Inference Layer. We search for $k=11$ neighbors for each point and divide them into batches with $N=1024$ points based on the partitioning method proposed in Section~\ref{Block Partitioning}. We converted the raw RGB attribute into the Y-CoCg space, and used the multi-channel attributes compression module in Section~\ref{crossChannel}. All three directions of the thresholds (omit $t_{-1}$) of DALD in Eqn.~\eqref{Eq2} are set to $\{0,1,3,\infty\}$. The parameters for LiDAR compression slightly change in the LoD structure because LiDAR usually has fewer points. The detailed parameters are shown in Table~\ref{Parameters}. We implement the deep entropy model and block partitioning by PyTorch, while the LoDs generation and arithmetic coding are implemented with the C++ backend. Our entropy model performs the training/testing on a computer with Inter Silver 4210 CPUs and a TITAN RTX GPU (24 G memory). We used batch sizes 32, epochs of 8, and an Adam optimizer with a $10^{-3}$ learning rate. It takes around two days to train each model. 

We compare our method with the state-of-the-art deep learning-based PCAC methods including 3CAC~\cite{wangS}, CNeT~\cite{CNET}, as well as its fast version MNet~\cite{MNet} for object point clouds and compare MuSCLE~\cite{MUSCLE} and 3CAC for LiDAR compression. We also compare our method with the popular hand-crafted method G-PCC with the latest version (TMC13v23), using the default setting (off angular mode) under CW conditions (lossless geometry and lossless attribute)~\cite{CTC} on the octree prediction-lifting branch. We use only the bits per point (Bpp) of the attribute to measure the compression ratio since this work aims at lossless attribute compression without any distortion.
 
\begin{table}[t]
\caption{Performance Comparison with G-PCCv23 over Bitrate (in Bpp) on MPEG CAT1}
\label{table1}
\setlength{\tabcolsep}{1.5mm}
\renewcommand\arraystretch{0.9415}
\begin{tabular}{c|l|ccc|c|c|cc}
\hline
                         & \makecell[c]{\ \ \ Point Cloud}   & Vox & \R{NN}    & \#Points & G-PCC & \R{$\rm bpp^{1}$}           &  \R{$\rm bpp^{2}$}            & Gain              \\ \hline\hline
                         & \cellcolor[HTML]{FFC7CE}{\color[HTML]{9C0006} queen\_0200}               & 10  & 42.55 & 1000993  & 7.65  & 6.93          & \textbf{6.83}  & -10.77\%          \\
                         & \cellcolor[HTML]{FFC7CE}{\color[HTML]{9C0006} basketball\_player\_00200} & 11  & 36.79 & 2925514  & 7.67  & \textbf{*}    & \textbf{6.43}  & -16.19\%          \\
                         & \cellcolor[HTML]{FFC7CE}{\color[HTML]{9C0006} facade\_00064}             & 11  & 36.30 & 4061755  & 10.11 & 9.66          & \textbf{9.60}  & -5.00\%           \\
                         & \cellcolor[HTML]{FFC7CE}{\color[HTML]{9C0006} dancer\_00000001}          & 11  & 36.05 & 2592758  & 7.74  & \textbf{*}    & \textbf{6.52}  & -15.78\%          \\
                         & \cellcolor[HTML]{FFC7CE}{\color[HTML]{9C0006} redandblack\_1550}         & 10  & 35.34 & 757691   & 9.18  & \textbf{8.15} & 8.32           & -9.38\%           \\
                         & \cellcolor[HTML]{FFC7CE}{\color[HTML]{9C0006} soldier\_0690}             & 10  & 35.07 & 1089091  & 6.92  & \textbf{*}    & \textbf{5.83}  & -15.81\%          \\
                         & \cellcolor[HTML]{FFC7CE}{\color[HTML]{9C0006} longdress\_1300}           & 10  & 34.75 & 857966   & 11.51 & \textbf{*}    & \textbf{10.24} & -11.06\%          \\
                         & \cellcolor[HTML]{FFC7CE}{\color[HTML]{9C0006} loot\_1200}                & 10  & 34.67 & 805285   & 6.06  & \textbf{5.11} & 5.24           & -13.51\%          \\
\multirow{-9}{*}{Solid}  & \cellcolor[HTML]{FFC7CE}{\color[HTML]{9C0006} thaidancer\_viewdep}       & 12  & 27.06 & 3130215  & 7.43  & \textbf{6.17} & 6.22           & -16.26\%          \\ \hline
                         & \cellcolor[HTML]{C6EFCE}{\color[HTML]{006100} house\_w/o\_roof\_00057}   & 12  & 10.29 & 4848745  & 10.25 & 9.52          & \textbf{9.47}  & -7.66\%           \\
                         & \cellcolor[HTML]{C6EFCE}{\color[HTML]{006100} head\_00039}               & 12  & 10.02 & 13903516 & 9.69  & 8.45          & \textbf{8.23}  & -15.08\%          \\
                         & \cellcolor[HTML]{C6EFCE}{\color[HTML]{006100} landscape\_00014}          & 14  & 9.65  & 71948094 & 11.06 & 10.53         & \textbf{10.18} & -7.95\%           \\
                         & \cellcolor[HTML]{C6EFCE}{\color[HTML]{006100} boxer\_viewdep}            & 12  & 8.87  & 3493085  & 5.92  & 4.90          & \textbf{4.85}  & -18.07\%          \\
                         & \cellcolor[HTML]{C6EFCE}{\color[HTML]{006100} loot\_viewdep}             & 12  & 8.82  & 3017285  & 4.96  & 3.84          & \textbf{3.77}  & -24.06\%          \\
                         & \cellcolor[HTML]{C6EFCE}{\color[HTML]{006100} longdress\_viewdep}        & 12  & 8.76  & 3096122  & 9.63  & 8.05          & \textbf{8.05}  & -16.44\%          \\
                         & \cellcolor[HTML]{C6EFCE}{\color[HTML]{006100} soldier\_viewdep}          & 12  & 8.75  & 4001754  & 6.13  & 4.78          & \textbf{4.77}  & -22.19\%          \\
                         & \cellcolor[HTML]{C6EFCE}{\color[HTML]{006100} redandblack\_viewdep}      & 12  & 8.70  & 2770567  & 7.63  & 6.54          & \textbf{6.48}  & -15.09\%          \\
                         & \cellcolor[HTML]{C6EFCE}{\color[HTML]{006100} frog\_00067}               & 12  & 5.85  & 3614251  & 9.67  & 9.10          & \textbf{8.95}  & -7.45\%           \\
                         & \cellcolor[HTML]{C6EFCE}{\color[HTML]{006100} facade\_00009}             & 12  & 3.63  & 1596085  & 11.57 & 10.55         & \textbf{10.35} & -10.55\%          \\
                         & \cellcolor[HTML]{C6EFCE}{\color[HTML]{006100} arco\_valentino\_dense}    & 12  & 3.15  & 1481746  & 15.73 & 14.99         & \textbf{14.78} & -6.04\%           \\
                         & \cellcolor[HTML]{C6EFCE}{\color[HTML]{006100} facade\_00015}             & 14  & 1.53  & 8907880  & 10.90 & 9.98          & \textbf{9.68}  & -11.23\%          \\
\multirow{-13}{*}{Dense} & \cellcolor[HTML]{C6EFCE}{\color[HTML]{006100} facade\_00064}             & 14  & 1.48  & 19702134 & 9.68  & 9.21          & \textbf{9.03}  & -6.70\%           \\ \hline
                         & \cellcolor[HTML]{FFEB9C}{\color[HTML]{9C5700} shiva\_00035}              & 12  & 1.37  & 1009132  & 15.23 & 14.63         & \textbf{14.34} & -5.81\%           \\
                         & \cellcolor[HTML]{FFEB9C}{\color[HTML]{9C5700} palazzo\_carignano\_dense} & 14  & 1.29  & 4187594  & 16.65 & 16.12         & \textbf{15.77} & -5.29\%           \\
                         & \cellcolor[HTML]{FFEB9C}{\color[HTML]{9C5700} stanford\_area\_2}         & 16  & 1.21  & 47062002 & 8.56  & 7.37          & \textbf{6.71}  & -21.63\%          \\
                         & \cellcolor[HTML]{FFEB9C}{\color[HTML]{9C5700} ulb\_unicorn}              & 13  & 1.17  & 1995189  & 6.94  & 6.06          & \textbf{5.77}  & -16.89\%          \\
                         & \cellcolor[HTML]{FFEB9C}{\color[HTML]{9C5700} staue\_klimt}              & 12  & 1.11  & 499660   & 13.40 & 12.13         & \textbf{11.93} & -10.96\%          \\
\multirow{-6}{*}{Sparse} & \cellcolor[HTML]{FFEB9C}{\color[HTML]{9C5700} egyptian\_mask}            & 12  & 1.01  & 272684   & 9.39  & 8.61          & \textbf{8.37}  & -10.90\%          \\ \hline
                         & \cellcolor[HTML]{DDEBF7}{\color[HTML]{2F75B5} ulb\_unicorn}              & 20  & 1.00  & 2000297  & 6.92  & 6.09          & \textbf{5.79}  & -16.34\%          \\
                         & \cellcolor[HTML]{DDEBF7}{\color[HTML]{2F75B5} palazzo\_carignano\_dense} & 20  & 1.00  & 4203962  & 16.63 & 16.10         & \textbf{15.74} & -5.37\%           \\
                         & \cellcolor[HTML]{DDEBF7}{\color[HTML]{2F75B5} facade\_00015}             & 20  & 1.00  & 8929532  & 10.87 & 9.99          & \textbf{9.77}  & -10.16\%          \\
                         & \cellcolor[HTML]{DDEBF7}{\color[HTML]{2F75B5} landscape\_00014}          & 20  & 1.00  & 72145549 & 10.79 & 10.53         & \textbf{10.21} & -5.46\%           \\
                         & \cellcolor[HTML]{DDEBF7}{\color[HTML]{2F75B5} arco\_valentino\_dense}    & 20  & 1.00  & 1530552  & 15.71 & 14.95         & \textbf{14.75} & -6.11\%           \\
                         & \cellcolor[HTML]{DDEBF7}{\color[HTML]{2F75B5} facade\_00064}             & 20  & 1.00  & 19714629 & 9.63  & 9.22          & \textbf{9.08}  & -5.76\%           \\
                         & \cellcolor[HTML]{DDEBF7}{\color[HTML]{2F75B5} head\_00039}               & 20  & 1.00  & 14025709 & 9.56  & 8.53          & \textbf{8.26}  & -13.58\%          \\
                         & \cellcolor[HTML]{DDEBF7}{\color[HTML]{2F75B5} egyptian\_mask}            & 20  & 1.00  & 272689   & 9.40  & 8.67          & \textbf{8.42}  & -10.43\%          \\
                         & \cellcolor[HTML]{DDEBF7}{\color[HTML]{2F75B5} facade\_00009}             & 20  & 1.00  & 1602990  & 11.48 & 10.79         & \textbf{10.40} & -9.38\%           \\
                         & \cellcolor[HTML]{DDEBF7}{\color[HTML]{2F75B5} frog\_00067}               & 20  & 1.00  & 3630907  & 9.58  & 9.23          & \textbf{8.99}  & -6.15\%           \\
                         & \cellcolor[HTML]{DDEBF7}{\color[HTML]{2F75B5} house\_w/o\_roof\_00057}   & 20  & 1.00  & 5001077  & 10.19 & 9.54          & \textbf{9.48}  & -7.01\%           \\
                         & \cellcolor[HTML]{DDEBF7}{\color[HTML]{2F75B5} shiva\_00035}              & 20  & 1.00  & 1010591  & 15.21 & 14.66         & \textbf{14.32} & -5.87\%           \\
\multirow{-13}{*}{Scant} & \cellcolor[HTML]{DDEBF7}{\color[HTML]{2F75B5} staue\_klimt}              & 20  & 1.00  & 499886   & 13.40 & 12.16         & \textbf{12.01} & -10.35\%          \\ \hline
  &\makecell[c]{\ \ \ \textbf{Average}}                                             & -   & -     & 8517004  & 10.16 & 9.51          & \textbf{9.12}  & \textbf{-11.36\%}\\\hline
\end{tabular}
    \begin{tablenotes}
      \item * The data is in the training set.
      \item ${}^1$ DALD-PCAC trained following CNeT; ${}^2$ DALD-PCAC trained on PCL-PCD dataset.
    \end{tablenotes}
\end{table}

\begin{table}[t]
\caption{Performance Comparison over Bitrate (in Bpp) with Other Deep-Learning Based Methods}
\setlength{\tabcolsep}{0.45mm}
\renewcommand\arraystretch{1}
\label{table3}
\begin{tabular}{c|c|c|c|c|ccccc}
\hline
Dataset                                                                       & Point Cloud          & Density        & \R{NN}         & \#Frames & G-PCC & MNeT   & CNeT  & 3CAC & DALD-PCAC \\\hline\hline
\multirow{6}{*}{\begin{tabular}[c]{@{}c@{}}MPEG\\CAT1 \\ vox12\end{tabular}}  & thai.\_viewdep         & Solid        & 27.06      & 1        & 7.43  & *          & 6.39      & 6.49    & \textbf{6.17}     \\ 
& soldier\_viewdep        & Dense       & 8.75          & 1        & 6.13  & *          & 5.63      & 5.55    & \textbf{4.78}     \\
& frog\_00067            & Dense       &5.85            & 1        & 9.67  & *          & 11.67     & *       & \textbf{9.10}     \\
& arco.\_dense           & Dense        &3.15          & 1        & 15.73 & *          & 17.00     & *       & \textbf{14.99}    \\
& shiva\_00035           & Sparse        &1.37         & 1        & 15.23 & *          & 20.75     & *       & \textbf{14.63}    \\
& staue\_klimt           & Sparse        &1.11         & 1        & 13.40 & *          & 22.79     & 12.80   & \textbf{12.13}    \\ \hline 
&   \textbf{Average}     &            &  -                     & -        & 11.26 & - 	         & 14.04     & - 	     & \textbf{10.30}    \\ \hline  \hline 
\multirow{2}{*}{\begin{tabular}[c]{@{}c@{}}8iVFBv2\\vox10\end{tabular}}  
& loot   & \multirow{2}{*}{Solid} & 34.28 & 300      & 6.08  & 8.38       & \textbf{4.65}      & 5.18    & 5.15     \\
& redandblack     &  & 35.02    & 300      & 9.15  & 12.23      & \textbf{7.21}      & 8.07    & 8.11     \\ \hline
\multirow{2}{*}{\begin{tabular}[c]{@{}c@{}}MVUB\\vox10\end{tabular}}   
& phil                   & \multirow{2}{*}{Solid} & 49.26 & 245      & 10.16 & 10.49      & \textbf{4.77}      & 6.78    & 6.88     \\
& ricardo          &               & 49.61    & 215      & 5.86  & 7.31       & \textbf{2.61}      & 3.59    & 3.56     \\ \hline
&    \textbf{Average}    &   &   -   &  -      & 7.81          & 9.60   & \textbf{4.81} & 5.91      & 5.93           \\\hline       
\end{tabular}
    \begin{tablenotes}
      \item * The performance is not comparable to G-PCC.
    \end{tablenotes}
\end{table}



\begin{table}[t]
  \centering
    \caption{Performance Comparison over Bitrate (in Bpp) with CNeT on Different Density Point Clouds}
    \setlength{\tabcolsep}{1mm}
    \renewcommand\arraystretch{1}
    \label{CNETDensity}
    \begin{tabular}{c|ccc|ccc}\hline
    Point   Cloud                      & \#Points & Density & NN    & G-PCC & CNeT  & DALD-PCAC \\\hline\hline
    house\_without\_roof\_00057\_vox12 & 4848745  & Dense   & 10.29 & 10.25 & 12.24 & \textbf{9.52}     \\
    house\_without\_roof\_00057\_vox20 & 5001077  & Scant   & 1.00  & 10.19 & 24.88 & \textbf{9.54}     \\
    facade\_00009\_vox12               & 1596085  & Dense   & 3.63  & 11.57 & 13.85 & \textbf{10.55}    \\
    facade\_00009\_vox20               & 1602990  & Scant   & 1.00  & 11.48 & 23.19 & \textbf{10.79}    \\
    egyptian\_mask\_vox12              & 272684   & Sparse  & 1.01  & 9.39  & 23.37 & \textbf{8.61}     \\
    egyptian\_mask\_vox20              & 272689   & Scant   & 1.00  & 9.40  & 23.26 & \textbf{8.67}     \\
    redandblack\_vox10\_1550           & 757691   & Solid   & 35.34 & 9.18  & \textbf{7.21}  & 8.15     \\
    redandblack\_viewdep\_vox12        & 2770567  & Dense   & 8.70  & 7.63  & 8.47  & \textbf{6.54}     \\
    loot\_vox10\_1200                  & 805285   & Solid   & 34.67 & 6.06  & \textbf{4.65}  & 5.11     \\
    loot\_viewdep\_vox12               & 3017285  & Dense   & 8.82  & 4.96  & 5.17  & \textbf{3.84}    \\
    \hline
    \textbf{Average}                   & 2094509  &  -      & 10.55 & 9.01  & 13.49 & \textbf{8.13}              \\\hline       
    \end{tabular}
\end{table}

\subsection{Experiment Results}
\subsubsection{Compared with GPCC} Table ~\ref{table1} shows the performance comparison of our method with G-PCC latest version 23. Our method is trained separately following CNeT~\cite{CNET} ($\text{bpp}^1$) and on PCL-PCD~\cite{pcd-pcl} dataset ($\text{bpp}^2$). DALD-PCAC achieves the best performance for all the point clouds, obtaining 9.23\% and 11.36\% bitrate gain over G-PCC on average respectively. Our model trained on PCL-PCD performs better, especially on sparse point clouds, probably because PCL-PCD is sparser than CNeT training set and has richer textures. The average bitrate saving of our model trained on PCL-PCD in Soild, Dense, Sparse and Scant categories over G-PCC is 12.64\%, 12.96\%, 11.91\% and 8.16\% (11.54\%, 11.53\%, 8.58\% and 6.33\% for $\text{bpp}^1$). The gain on denser point clouds appears to be higher, which may be due to the fact that deep learning-based methods are easier (compared to traditional methods) to learn correlations on denser point clouds. 
\subsubsection{Compared with Deep-Learning Methods} Table~\ref{table3} shows the performance comparison with other deep-learning based methods. \R{Our method is trained on CNeT training set to make a fair compassion.} Our approach maintains advanced performance across all of the sparse point clouds (\ie, MPEG CAT1 vox12). Our method also has a good performance on dense data. The data in MVUB~\cite{MVUB} is even denser than the Solid categories of MPEG CAT1 and 8iVFBv2. For those Solid point clouds, our method has a comparable performance with the latest 3CAC~\cite{wangS} and has a faster speed (see Table~\ref{Time}). CNeT~\cite{CNET} achieves best in the Solid point clouds, but its decoding speed is unacceptably slow because the voxel-wise autoregressive design. Although our method is not as good as CNeT~\cite{CNET} on those point clouds, our method is much better than its fast version MNeT~\cite{MNet}, which is a multi-scale parallel version of CNeT with a similar complexity to our method. These experiment results demonstrate the excellent performance and robustness of our method on most point clouds.
\subsubsection{Compared with CNeT on PC Pairs} Table \ref{CNETDensity} shows the comparison between our method and CNeT on point cloud pairs with different resolutions in MPEG CAT1. These point cloud pairs have similar content except for resolution and density. The performance of convolution method is degenerated when the density (\ie, NN) decreases, especially when the density changes from Dense to Scant, \eg, \textit{house\_without\_roof} and \textit{facade\_00009}. This confirms our observation that the performance of the convolution-based methods is greatly affected by the density. While our method outperforms G-PCC 9.7\% and CNeT 39.7\% on average, and exhibits a clear advantage over others in all the non-solid point clouds. Since LoD and DALD which can embed a fixed number of neighbors, our approach is robust to the density variations, with trends similar to G-PCC performance changes.
\subsubsection{LiDAR Compression}
Table \ref{table4} shows our performance in LiDAR compression. Our method outperforms others and achieves a 15\% gain over G-PCC on average. Compared with the latest sparse convolution-based method 3CAC~\cite{wangS}, our method can save 7.9\% bitrates and has much faster encoding and decoding speed. The convolution-based methods are challenging to obtain sufficient context in a sparse scene such as LiDAR, which limits their performance, and the cross-group or point-wise autoregressive technique used to improve the performance may increase the complexity. 

\begin{table}[t]
\centering
\caption{Performance comparisons over Bitrate (in Bpp) and Runtime (in Seconds) on the LiDAR}
\label{table4}
\setlength{\tabcolsep}{0.8mm}
\renewcommand\arraystretch{1}
\begin{threeparttable}
\begin{tabular}{c|c|c|c|cccc}
\hline 
LiDAR                        	& Resolution &NN &\#Frames 	& G-PCC     & 3CAC~\cite{wangS}   & MuSLCE~\cite{MUSCLE}  & DALD-PCAC      \\\hline \hline
\multirow{2}{*}{Ford 02-03}     & 1 mm  & 1.49	&1500	 & 5.22      & 4.97     & *       &  \textbf{4.52}      \\
						      & 2 cm  & 1.94  &1500    & 5.15      & 5.00     & *       &  \textbf{4.65}        \\
\hline
\multirow{3}{*}{KITTI 11-21} 	& 12 bits & 22.93  &20351  & 2.38   	& *        & 2.13    & \textbf{2.05}         \\
							    & 14 bits & 7.46  &20351 &  4.49  	    & *        & 4.30    & \textbf{3.72}       \\
							    & 16 bits & 1.42 &20351  &  4.80       & *        & 4.68    & \textbf{3.94}       \\
\hline\hline
\multirow{2}{*}{Ford\_q1mm}     &\multicolumn{3}{c|}{Enc. Time (s/frame)}   &\textbf{0.9} & 8.0      & *       & 1.7   \\                 
    &\multicolumn{3}{c|}{Dec. Time (s/frame)}   &\textbf{0.9} & 8.0      & *       & 1.8     \\
\hline
\end{tabular}
\end{threeparttable}
\begin{tablenotes}
  \item * Data is missing because the code is unavailable.
\end{tablenotes}
\end{table}

\subsection{Complexity Analysis}
\label{Complexity Analysis}
Table~\ref{Time} shows that our method has low complexity. It has the fastest encoding/decoding speed among the deep learning-based PCAC methods and achieves end-to-end coding as fast as G-PCC, \R{which mainly because our method abandons intra-prediction in inference, thus avoiding the autoregressive problem in the other learning-based methods. Meanwhile, due to the shallow network (only three layers of Transformer)}, the inference speed of our deep entropy model is relatively fast, which only takes about 6\% in the whole process. Hence, we believe there are many opportunities to speed up our method further, such as polishing the implementation of LoDs generation and arithmetic coding or interleaving them with GPU tasks. Our model size is also sufficiently small, which only has 0.63 M parameters with 9.6 MB of disk storage and requires less than 3 GB of GPU memory for inference, while 3CAC~\cite{wangS} has about 18 M parameters, and CNeT~\cite{CNET} contains three checkpoints, each with 1 GB of disk. The lightweight architecture makes our model easier to run on mobile devices such as autonomous vehicles and VR equipment for practical applications. 

\begin{table}[t]
\caption{Model Size and Runtime Comparison (tested on 8iVFBv2 vox10) }
\begin{tabular}{c|ccccc}\hline
Method                  & MNeT & CNeT    & 3CAC  & DALD-PCAC \\\hline\hline
Model Size \R{(disk storage)}             & 52MB & 960MB*3 & 201MB & \textbf{9.6MB}    \\
Encode (s/frame)          & 22.01    & 42.3    & 15.70  & \textbf{9.68}     \\
Decode (s/frame)          & 29.18   & 1528    & 16.00  & \textbf{11.26}   \\\hline
\end{tabular}
\label{Time}
\end{table}


\begin{figure}[t]
\centering
{
\subfigure[]{
\includegraphics[width=.34\linewidth]{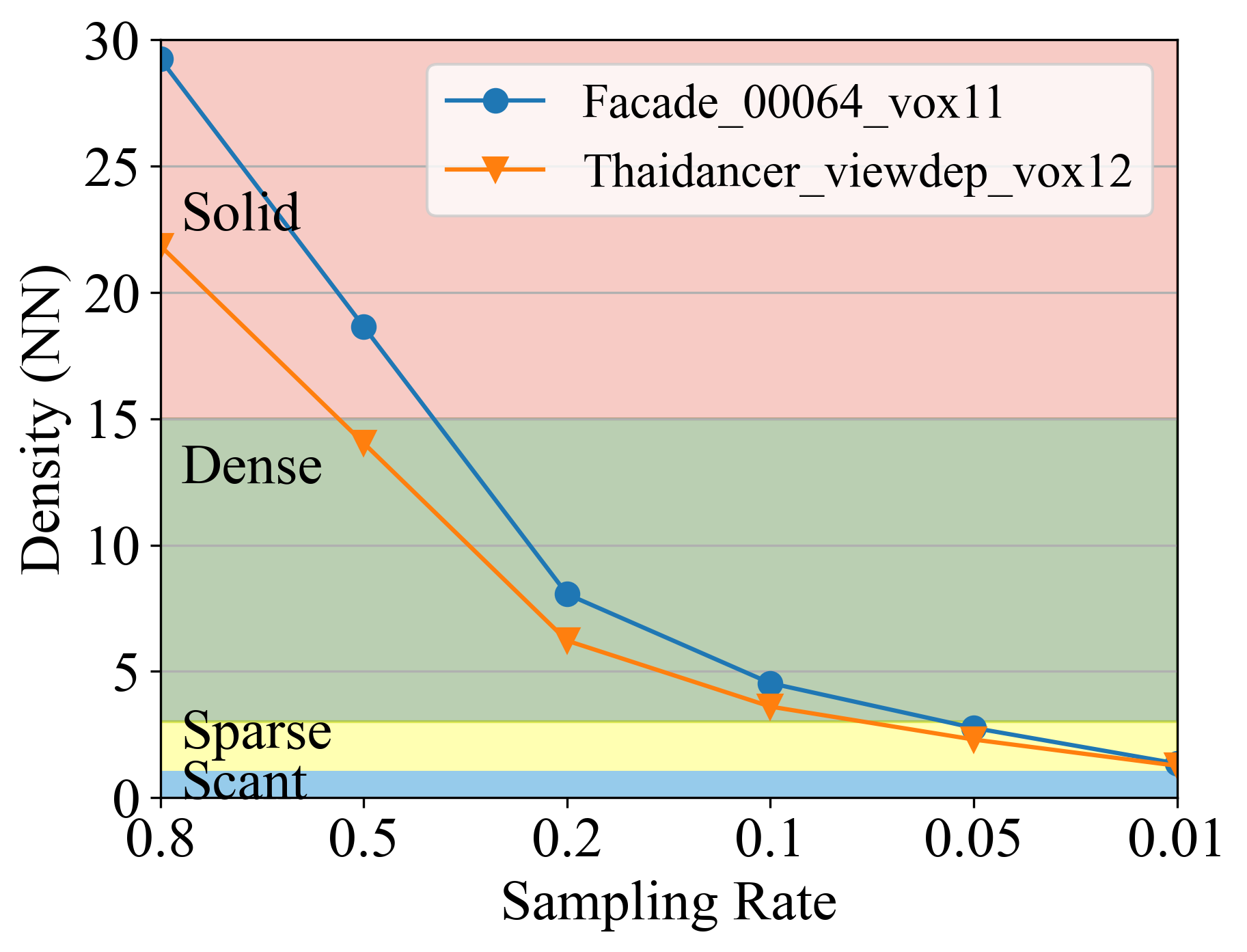}
} 
\subfigure[]{
\includegraphics[width=.625\linewidth]{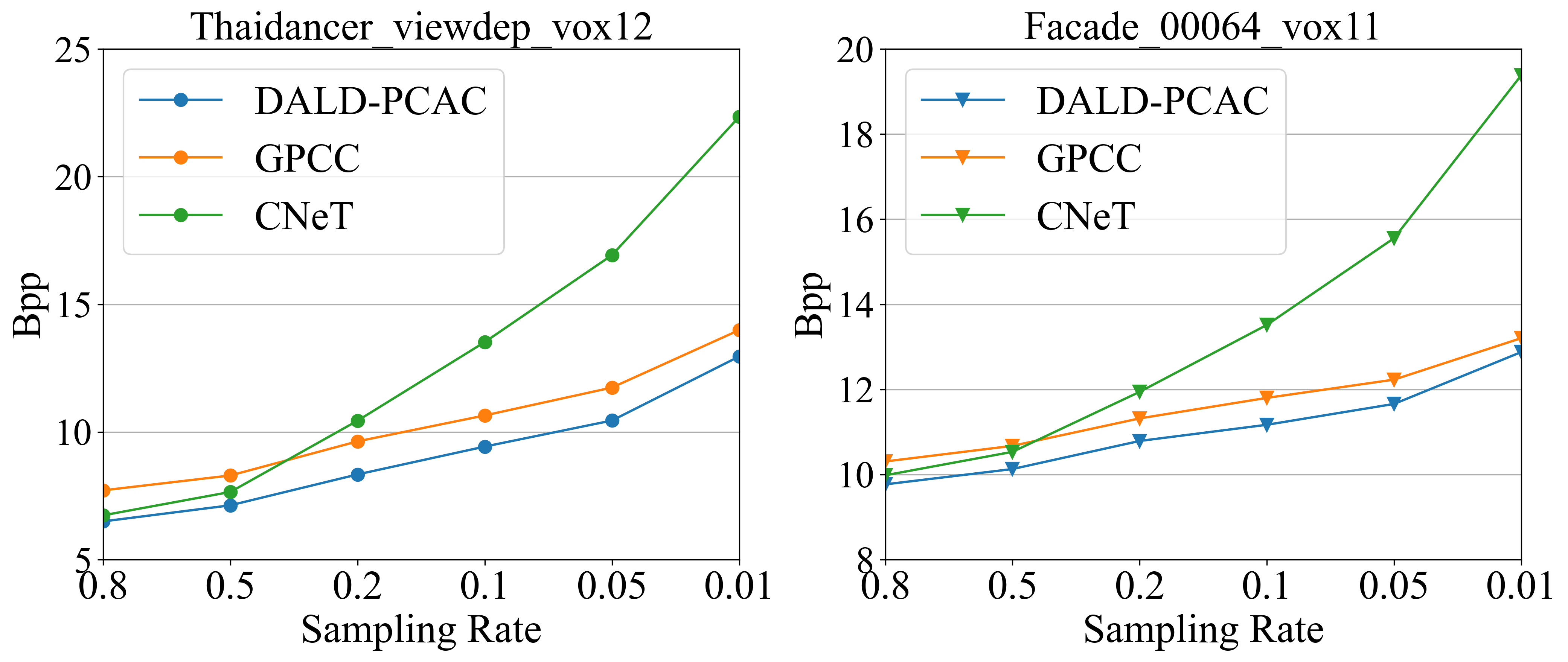}
}
\caption{\R{(a) Density (\ie, average number of neighbors in a $5\times5\times5$ convolution kernel) versus different sampling ratios. (b) Rate comparison on point clouds with different sampling ratios. }}
\label{Robust}
}
\end{figure}

 \begin{figure}[t]
\centering
{
\subfigure[]{
\includegraphics[width=.4\linewidth]{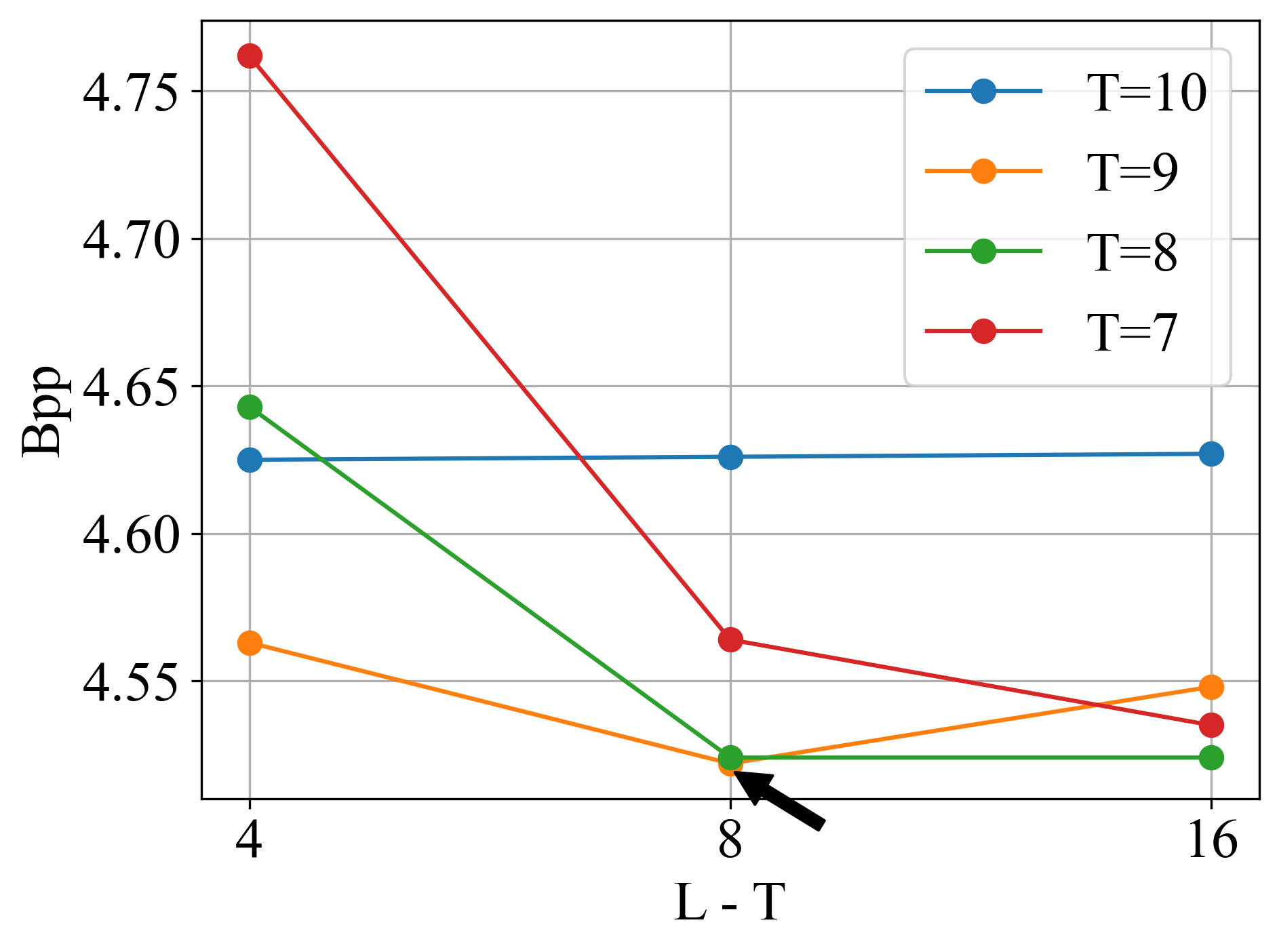}
}\hspace{10mm}
\subfigure[]{
\includegraphics[width=.4\linewidth]{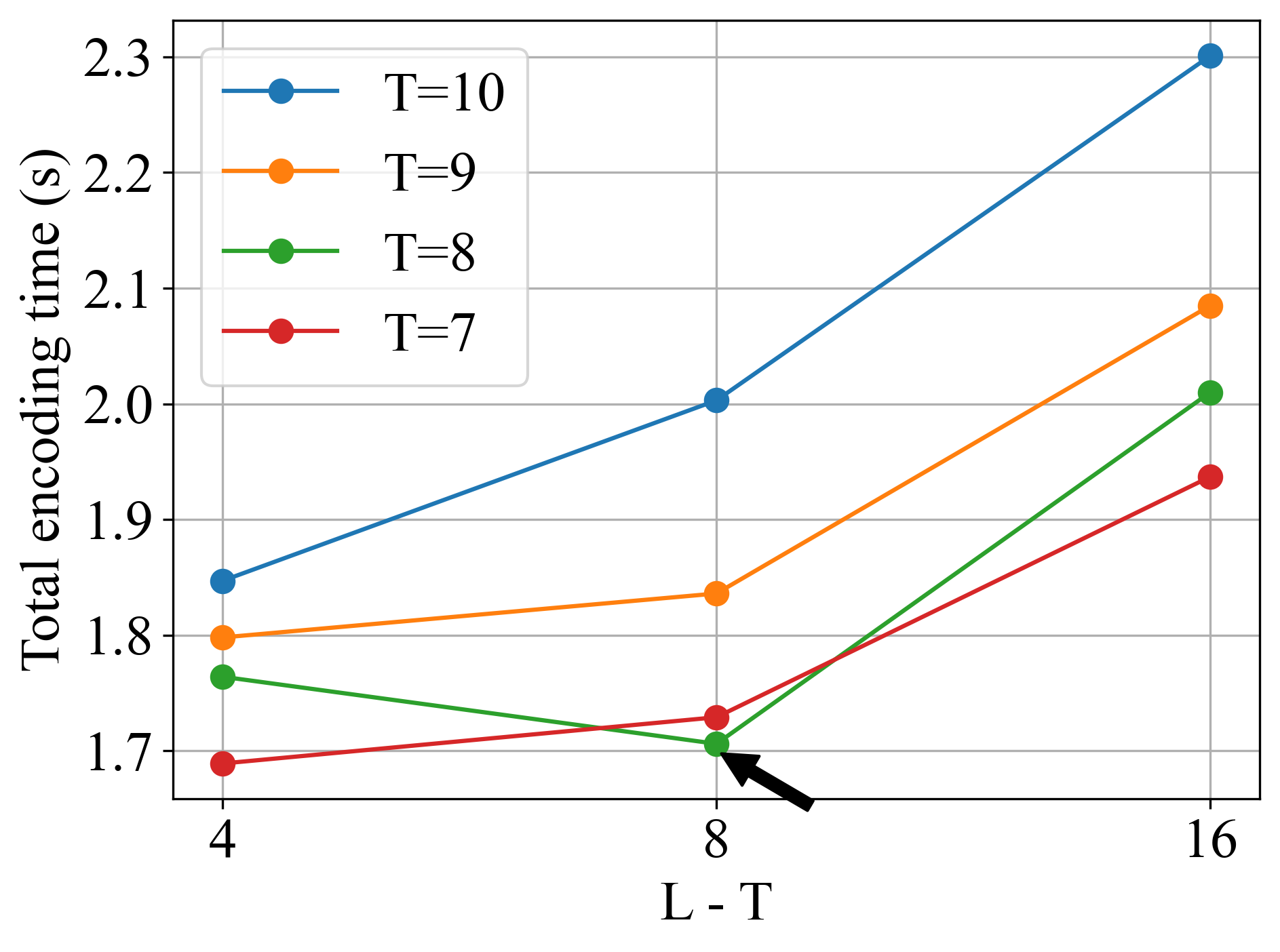}
}
\caption{Ablation study of LoD structure. (a) Bpp; (b) Encoding time.}
\label{LoD_ab}
}
\end{figure}

\subsection{\R{Parameter Analysis and Ablation Study}}
\subsubsection{Robustness to density}
We performed additional experiments to validate the robustness of DALD-PCAC to
variation in density. We test G-PCC, CNeT~\cite{CNET}, and our method on \R{\textit{Thaidancer\_viewdep\_vox12} and \textit{Facade\_000064\_vox11}} without retraining. The two PCs are randomly sampled by ratios of 0.8, 0.5, 0.2, 0.1, 0.05 and 0.01 and then perform lossless attribute compression. As shown in Figure~\ref{Robust} (a), the density of the data decreases as the sampling ratio decreases. From Figure~\ref{Robust} (b), \R{we notice that our method retain about $12\%$ and $5\%$ gain over G-PCC respectively after density of Solid,} while the performance of CNeT deteriorates due to the sparsity, especially when the density decreases after Solid. These results validate the density robustness of the proposed Density-Adaptive Local Descriptor, which makes DALD-PCAC can be applied to point clouds with varying densities.

\subsubsection{Detailed Setting of LoD}
The LoD structure determines the number of points encoded in the Base Layer and Inference Layer. Since we use different compression methods (\ie, run-length coding and deep entropy models) and autoregressive design (\ie, intra-layer and inter-layer) in these two layers, we need to examine the impact of the LoD structure on performance carefully. Figure~\ref{LoD_ab} shows the results of the ablation study of the LoD structure on the Ford\_q1mm dataset, where $T$ represents the number of refinement layers in the Base Layer and controls the proportion of the points of the two layers. $L-T$ is the number of refinement layers in the Inference Layer, which affects the number of autoregressive stages. As shown in Figure~\ref{LoD_ab} (a), we can increase the number of autoregressive stages (\ie, increase $L-T$) to obtain a lower bitrate, except that the proportion of the Base Layer is large (\ie, $T=10$). When $T=10$, as shown in Eqn.~\eqref{Eq14}, the Base Layer points amount $N_\text{base}$ accounts for 30\%, which occupies 40\% bitrate, resulting in the deep entropy model compresses very few points, leading to little change in the Bpp. As shown in Figure~\ref{LoD_ab} (b), the encoding time increases with the increase of $L-T$ because the serial autoregressive coding also reduces the parallelism. We choose $T=8$ and $L=16$ for LiDAR compression to maintain the trade-off between the coding time and the \R{compression} performance. With this setup, the Base Layer contains 5.4\% of the points and consumes 8.1\% of the bitrate, which means that the main contribution comes from the deep entropy model.

 \begin{figure}[tb]
\centering
{
\subfigure[]{
\includegraphics[width=.38\linewidth]{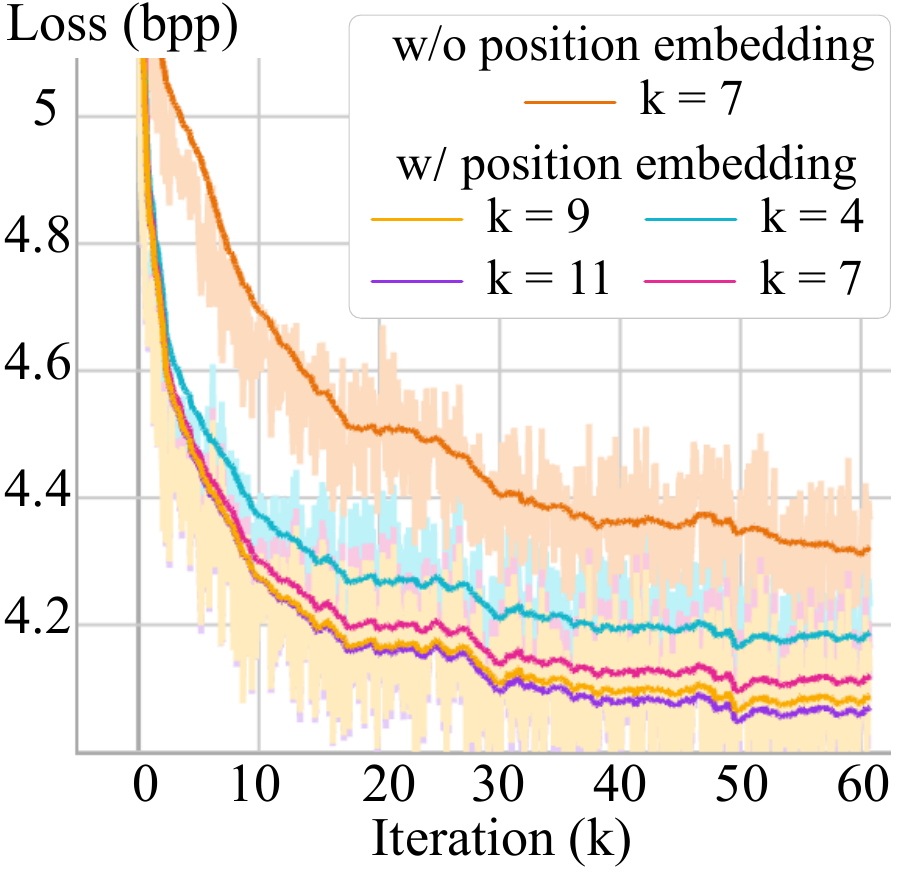}
}\hspace{10mm}
\subfigure[]{
\includegraphics[width=.38\linewidth]{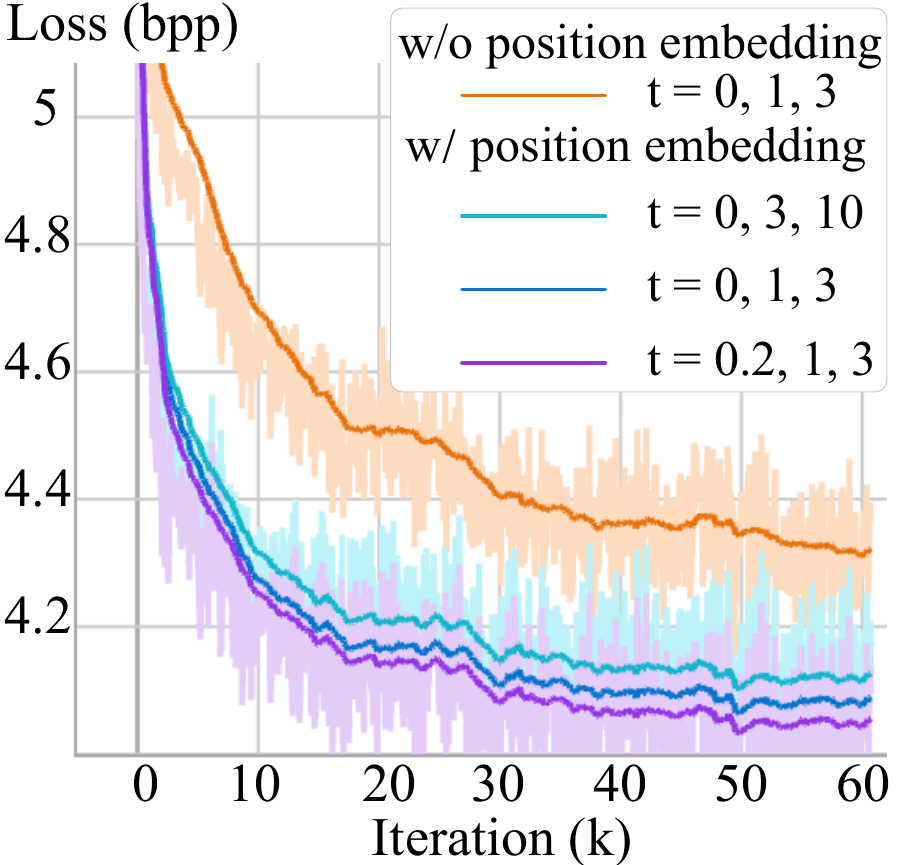}
}
\caption{Learning curves in ablation study with/without relative position embedding. (a) Different number of nearest neighbors $k$; (b) DALD thresholds $\{t_i\}$ in Eqn.~\eqref{Eq2}.}
\label{lossk}
}
\end{figure}

\begin{table}[t]
\centering
\caption{Ablation Study on the Thresholds in Relative Position Embedding.}
\label{thresholds}
\renewcommand\arraystretch{1}
\begin{tabular}{ccc|ccc}
\hline
\multicolumn{3}{c|}{$\{t^x_i\}=\{t^y_i\}=\{t^z_i\}=\{t_i\}$}                           & \multicolumn{3}{c}{$\{t^x_i\}=\{t^y_i\}=\{0.2, 1, 3, \infty\}$}                            \\ 
\multicolumn{1}{c}{$\{t_i\}$} & \#bins & bpp            & \multicolumn{1}{c}{$\{t^{z}_i\}$} & \#bins & bpp            \\ 
\hline\hline
MLP                  & 0       & 4.790          & 0,3,$\infty$           & 147     & 4.611          \\
1,3,5,$\infty$                & 343     & 4.642          & 1,3,5,$\infty$         & 343     & 4.599          \\
0,1,3,5,$\infty$              & 729     & 4.599          & 0,1,3,$\infty$         & 343     & 4.592          \\
0.2,1,3,$\infty$              & 343     & \textbf{4.580} & 0.2,0.4,1,$\infty$     & 343     & \textbf{4.533} \\ 
\hline
\end{tabular}
\end{table}

\subsubsection{Block Partitioning}
To validate our proposed block partitioning method, we replace the block partitioning method with KD-tree based partitioning. We retrain the model on object point clouds with the same settings. Experiments show that the block partitioning method based on base-layer priors can provide more consistent results with lower standard deviation and save approximately 1.8\% bitrate on average. Moreover, the original method based on KD-tree has a higher bitrate even if at the same deviation, which implies that the inconsistent blocks may adversely affect the generalization and performance of the deep entropy models.

\subsubsection{Number of Nearest Neighbors} 
We search a different number of nearest neighbors $k$ in the prediction and construction of the density-adaptive learning descriptor (DALD). We set $\{t_i\}=\{0,1,3,\infty\}$ and evaluate the deep entropy model with $k={4,7,9,11}$. The learning loss in the Ford dataset is shown in Figure \ref{lossk} (a). It shows that involving more neighbors improves performance in general. Thus, we choose $k=9$ as the default setting for the trade-off between the marginal effect of bitrate and computational complexity in LiDAR compression.

\subsubsection{Relative Position Embedding}
\R{Relative position embedding is core component of neighbors embedding in DALD, which introduced in Eqn.~\eqref{Eq2}}. We tested its effectiveness on the Ford\_q1mm dataset with $k=7$. The ablation study on the thresholds of DALD is shown in Figure~\ref{lossk} (b), which shows that relative position embedding can significantly reduce the learning loss of the entropy model. The quantitative results are shown in Table \ref{thresholds}, where ``MLP'' means that we replace the function $E_l(\cdot)$ in Eqn.~(\ref{Eq6}) by multi-layer perceptron with a similar number of parameters and ``\#bins'' represents the number of labels in Eqn.~(\ref{eq3}). Table \ref{thresholds} shows that the relative position embedding provides more than 5.3\% gain over the ``MLP'' setting. Assigning more labels to smaller regions can achieve lower bitrates, which implies that allocating more learnable weights to closer neighbors with stronger correlations is more beneficial. Meanwhile, assigning more labels also helps to improve the performance of relative position embedding. 



\begin{figure}[t]
\centering
{\includegraphics[scale=0.65]{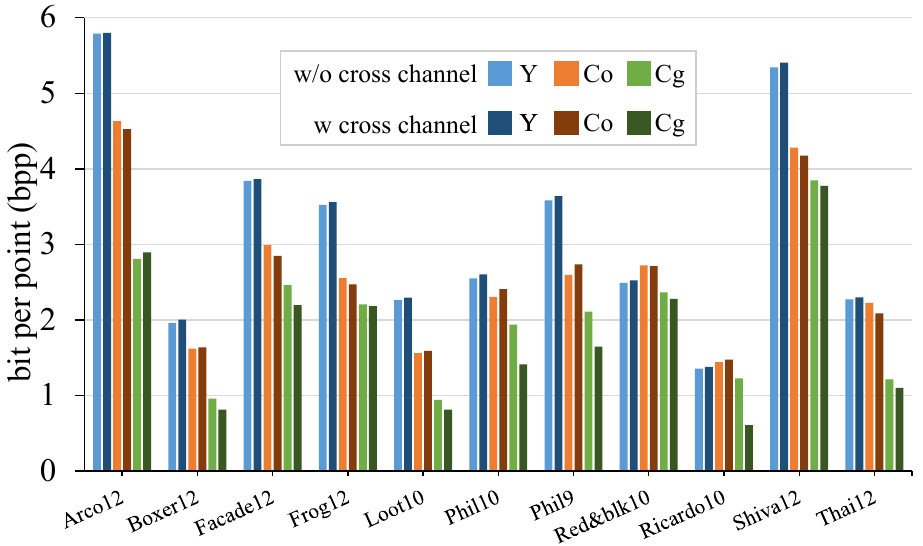}
\caption{Ablation study on bit allocated for each attribute.}
\label{crossYUV1}
}
\end{figure}


\subsubsection{Multi-Attributes Compression}
In the multi-attributes compression, we convert the RGB colors into the YCoCg-R space to remove the correlation between channels, which is validated in many works~\cite{CNET,wangS,Deep-PCAC}. The bit allocated for each attribute for some samples is shown in Figure~\ref{crossYUV1}, where ``w/o inter-channel'' means that we remove the prior information $r_{Y_i}$ and $r_{U_i}$ in Eqn.~\eqref{eq10} and encode the three channels separately. On average, the inter-channel technique we proposed in section~\ref{crossChannel} can save 2.6\% bitrate, similar to 3CAC~\cite{wangS}. Generally speaking, the inter-channel technique requires more bits on the Y component but saves more bits on the Co and Cg components. Most point clouds need more bits to encode the Y component than the Co and Cg components, but some point clouds like \textit{Redandblack} and \textit{Thaidancer} are the exceptions, \R{which probably because they have brighter color and higher chrominance}. A similar case is also observed in CNeT~\cite{CNET}.

\section{Conclusion and future work}
This paper presents a learned framework based on the levels of detail (LoD) structure for lossless attribute compression. We introduce a novel density-adaptive learning descriptor (DALD) designed to discover intricate structures and aggregate context information. Our approach, DALD-PCAC, not only achieves state-of-the-art performance on LiDAR and most point clouds but also has \R{lower computation complexity} compared to existing methods. A potential limitation of this work is that only lossless compression is discussed in this paper. \R{In future works, the density-adaptive learning descriptor can be extended to applications including lossy compression, mesh coding and point cloud processing, which aims to obtain sufficient context in point clouds of varying densities. By adopting the proposed descriptor to these areas, point context mining and modeling can be improved, ultimately improving the overall performance of these applications.}

\begin{acks}
  The research was partially supported by the RGC General Research Fund 11200323,  NSFC/RGC JRS Project N\_CityU198/24, and Hong Kong Innovation and Technology Fund GHP/044/21SZ, and PRP/036/24FX.
\end{acks}

\bibliography{ref}

\appendix

\end{document}